\documentclass[a4paper,11pt]{article}
\usepackage{jcappub} 
\usepackage{aas_macros} 
\usepackage{lineno}

\usepackage{graphicx}	
\usepackage{amsmath}	

\def\bi#1{\hbox{\boldmath{$#1$}}}

\usepackage[normalem]{ulem}

\title{\boldmath Enhancing CMB map reconstruction and power spectrum estimation with convolutional neural networks}

 \author[a,b,1]{Bel\'en Costanza\note{Corresponding author.},}
 \author[a,b]{Claudia G. Sc\'occola}
 \author[c]{and Mat{\'\i}as Zaldarriaga}
 \affiliation[a]{Facultad de Ciencias Astron\'omicas y Geof\'isicas, Universidad Nacional de La Plata, Observatorio Astron\'omico, Paseo del Bosque,  B1900FWA La Plata, Argentina}
 \affiliation[b]{Consejo Nacional de Investigaciones Cient\'ificas y T\'ecnicas (CONICET), Rivadavia 1917, Buenos Aires, Argentina}
 \affiliation[c]{School of Natural Sciences, Institute for Advanced Study, 1 Einstein Drive, Princeton, NJ 08540, USA}

\emailAdd{belen@fcaglp.unlp.edu.ar}

\abstract{The accurate reconstruction of Cosmic Microwave Background (CMB) maps and the measurement of its power spectrum are
crucial for studying the early universe. In this paper, we implement a convolutional neural network to apply the Wiener Filter to CMB temperature maps, and use it intensively to compute an optimal quadratic estimation of the power spectrum. 
Our neural network has a UNet architecture as that implemented in WienerNet, but with novel aspects such as being written in \textsc{python~3} and \textsc{TensorFlow~2}. It also includes an extra channel for the noise variance map, to account for inhomogeneous noise, and a channel for the mask.
The network is very efficient, overcoming the bottleneck that is typically found in standard methods to compute the Wiener Filter, such as those that apply the conjugate gradient. It scales efficiently with the size of the map, making it a useful tool to include in CMB data analysis. 
The accuracy of the Wiener Filter reconstruction is satisfactory, as compared with the standard method. We heavily use this approach to efficiently estimate the power spectrum, by performing a simulation-based analysis of the optimal quadratic estimator. We further
evaluate the quality of the reconstructed maps in terms of the power spectrum and find that we can properly recover the statistical properties of the signal.
We find that the proposed architecture can account for inhomogeneous noise efficiently. Furthermore, increasing the complexity of the variance map presents a more significant challenge for the convergence of the network than the noise level does.
}

\begin{document}

\maketitle

\flushbottom

\section{Introduction}
\label{sec:intro}

The cosmic microwave background (CMB) is the relic radiation from the early Universe released at recombination, 380,000 years after the Big Bang. It is one of the richest sources of information about the origin and evolution of our Universe. 
The past decades have witnessed a steady increase in the amount of CMB data, both in temperature and polarization~\citep{Planck2020,ACT2020a, ACT2020b,SPTPol2018}.
Sophisticated tools are required to extract accurate and precise cosmological information, thereby maximizing the scientific returns. 

The data analysis process in an experiment involves three essential data compression stages: the conversion of raw time-ordered data (TOD) into a sky map (map making), the estimation of power spectra, and the derivation of cosmological parameters from the power spectra. If all three data compression steps are lossless, the data pipeline will yield parameter measurements with error bars as small as those obtained through likelihood analysis directly from the TOD~\citep{1997PhRvD..55.5895T}. 
The Wiener filter~\citep{Wiener1949} is a powerful method for reconstructing signals by leveraging statistical knowledge of the signal and noise properties.
Wiener filtering of the maps is essential for achieving statistically optimal results in CMB data analysis, particularly when data compression involves any loss. However, the computational demands for deriving the exact Wiener filter solution have increased significantly due to the vast amount of data involved.

In this paper, we address the challenge of optimal map reconstruction and power spectrum estimation using noisy and sparsely sampled data. For the CMB, where the data undergoes a linear transformation of initial modes, the Wiener Filter (WF) emerges as an optimal filter for reducing noise in both simulated and real maps, resulting in an optimal signal reconstruction. The WF is a powerful approach that minimizes residual variances when reconstructing the true underlying field from noisy observations. Moreover, in the case of Gaussian-distributed data, the minimum variance estimator aligns with the maximum posterior probability estimator~\citep{Seljak1998}. 
The power spectrum is commonly used to extract cosmological information from CMB maps. We will take advantage of the efficient computation of the Wiener-filtered maps obtained with the neural networks, and implement the algorithm suggested by~\cite{Seljak2017}, in which an optimal quadratic estimator of the power spectrum is obtained from the maximization of the likelihood. The algorithm is efficient, thanks to the fast evaluation of the different terms that appear in the estimator, like the Hessian matrix, using simulations.

While the WF is the optimal filter for map reconstruction, it can be computationally expensive. The WF matrix, which represents the minimum variance estimator, involves inverting the covariance matrix of both the noise and signal. Typically, the noise covariance is diagonal in configuration space, and the signal covariance is diagonal in Fourier space. However, the combined covariance matrix is not diagonal (or sparse) on any basis, making the inverse covariance matrix computationally expensive to obtain, particularly for large datasets.

There are alternative methods used to solve the WF map such as the use of a preconditioner in the conjugate gradient method~\citep{10.5555/1403886} for matrix inversion, since in CMB problems matrices are often ill-conditioned~\citep{2004ApJS..155..227E}. There are several preconditioners proposed for CMB problems, such as the inverse of the diagonal part of the matrix in the linear system equation to solve~\citep{2004PhRvD..70h3511W} or the multigrid preconditioner~\citep{Smith2007}. Moreover, finding an efficient preconditioner is a complicated task, as it should provide a sufficiently accurate approximation to the true inverse while maintaining sparsity~\citep{1999ApJ...510..551O}. 

Other methods implement the WF without preconditioners. Instead, they utilize an iterative reconstruction process that incorporates a messenger field, facilitating communication between bases where signal and noise properties can be conveniently specified~\citep{Elsner2013,Kodi2017}. A different method without preconditioners involves optimization problems~\citep{Seljak2017,Horowitz2019} solving the minimization of a nonlinear function $\chi^{2}$ in order to maximize the posterior of the signal given the data.

In this study, we explore a machine learning-based approach for simulating the Wiener Filter using a neural network known as WienerNet, as proposed by~\cite{Munchmeyer2019}. Our investigation focuses on evaluating the performance and efficiency of the WF implemented with WienerNet under various noise models and CMB maps, considering different pixel counts and the flat sky approximation. We extend the original neural network, which was initially designed for homogeneous noise, to accommodate inhomogeneous 
noise. To achieve this, we incorporate an additional channel that captures the information about the inhomogeneous noise into the original neural network architecture. Moreover, we have written our codes in \textsc{tensorflow~2} and \textsc{keras}, which are based on \textsc{python~3}, as opposed to the original WienerNet, which was coded in \textsc{tensorflow~1}, and  \textsc{python~2.7}. 
Furthermore, we compare the outcomes obtained with the neural network 
to the results generated by the standard method of implementing the WF using the conjugate gradient approach, employing the \textsc{NIFTy} software\footnote{\url{https://ift.pages.mpcdf.de/nifty/}} as a benchmark.

In addition to the estimation of the field, we also present the estimation of the power spectrum amplitudes and their covariance matrix. We followed the proposal by~\citep{Seljak2017} which assumes a flat prior on the bandpowers and performs the maximum likelihood method after marginalization over the field, as opposed to~\citep{2016MNRAS.455.4452A}, where they jointly sample the posterior distribution of the field and the power spectra, in the Cosmic Shear scenario. In the linear regime, the power spectrum can be estimated by the optimal quadratic estimator~\citep{1997PhRvD..55.5895T}, which is quadratic on the data, after the optimal field reconstruction. This method is often replaced by the pseudo-$C_{\ell}$ method~\citep{2001PhRvD..64h3003W} assuming that the covariance matrix of the data is diagonal, which is not an optimal solution, as the correct weighting of the data involves multiplying it by the inverse of the covariance matrix. Here, we address the computational cost of weighting data with the Wiener Filter using neural networks. 

In this work, we present a first step towards addressing the problem of power spectrum estimation of incomplete, noisy CMB  maps. In this paper, we focus on temperature maps, but our long-term goal is to provide a framework that will be useful for analyzing future CMB polarization data, for which the complexity of the noise properties complicates the application of standard tools. One of these experiments is the Q \& U Bolometric Interferometer for Cosmology (QUBIC)~\cite{QUBIC_universe}. The multipeak synthesized beam of this telescope allows to make spectro-imaging~\cite{QUBIC_II} within the frequency band, at the cost of having spatial and subfrequency correlations. The implementation of machine learning tools in the analysis of QUBIC data is one of the main drivers for these investigations.  

The paper is organized as follows:  In section~\ref{sec:Background}, we briefly present the basics of Wiener Filtering and the power spectrum estimator we will use in this work. In section~\ref{sec:CNN} we described general Machine Learning concepts that are used to build a neural network and the WienerNet architecture. In section~\ref{sec:datasets}, we explain how we simulate our data, with different noise properties. We also create maps from a fiducial power spectrum, that are needed to estimate the power spectrum of the reconstructed signal.
Then, we present the results of applying WienerNet to CMB temperature maps with homogeneous noise (in section~\ref{sec:results_hom}) and inhomogeneous noise (in section~\ref{sec:results_inhom}). We compare the performance of the neural network approach, in comparison with the exact method, and evaluate the computational efficiency of the new approach. In addition, we present our results for power spectrum reconstruction, for the homogeneous case, in section~\ref{sec:results_hom}, and for the inhomogeneous case, in section~\ref{sec:results_inhom}.
The discussion and conclusions are presented in section~\ref{sec:conclusion}.

\section{Background}
\label{sec:Background}

In this section, we introduce the methods and tools used in this paper. Firstly, we present the Wiener filter, which we will apply to flat CMB noisy maps. Additionally, we provide an overview of an algorithm used to estimate the power spectra of the CMB temperature maps after applying the Wiener filter.

\subsection{Wiener Filter}

The Wiener Filter is a signal processing technique used to enhance signal quality by reducing noise and interference in the presence of random disturbances. In the context of CMB temperature maps, it improves the signal-to-noise ratio, extracting valuable cosmological insights from observations potentially affected by noise and uncertainty. In the case of masked data, it also reconstructs the large-scale signal inside the masked region.

Let us assume that the measurements $d$ of an underlying field $s$ that we want to estimate is a linear combination of the field, where $R$ is the 
response matrix of the measurement procedure and $\epsilon$ is the data uncertainty:
\begin{equation}
    \textbf{d} = \textbf{R} \textbf{s} + \epsilon
    \label{eq:eq1}
\end{equation}

In general, the application of a filter can be mathematically represented as a convolution, establishing a linear connection between the reconstructed underlying field ($y$), which represents the filtered map, and the input data ($d$), as expressed by the equation:
\begin{equation}
y = Md.
\label{eq:M}
\end{equation}
For the exact Wiener filter, the matrix $M$ is determined by minimizing the variance of the residual between the reconstructed underlying signal and the original signal, yielding the expression for the Wiener filter matrix, $M$:
\begin{equation}
\textbf{M} = \textbf{S}(\textbf{S}+\textbf{N})^{-1}\textbf{R}^{-1},
\label{eq:eq10}
\end{equation}
where $\textbf{S}$ and $\textbf{N}$ are the covariance matrices of the signal and noise, respectively, and $\textbf{R}$ is the response matrix of the measurement procedure. In the specific case where the underlying signal is a Gaussian random field, the Wiener filter estimator obtained by minimizing the variance of the residuals coincides with the Bayesian estimator that maximizes the conditional probability of the signal given the data~\citep{Zaroubi1995}:

\begin{equation}
 P(\textbf{s} |\textbf{d})  \propto 
     \exp \left[-\frac{1}{2}(\textbf{s}^{\dag}\textbf{S}^{-1}\textbf{s}+ (\textbf{d}-\textbf{R}\textbf{s})^{\dag}\textbf{N}^{-1}(\textbf{d}-\textbf{R}\textbf{s}))\right].
   \label{eq:eq16}                     
\end{equation}
The Wiener Filter estimator is, therefore, the optimal reconstruction of the signal as it represents the most probable configuration of the field given the data.

We perform the exact Wiener Filter through the \textsc{NIFTy} library~\citep{2013A&A...554A..26S}, which is a software designed to enable the development of signal inference algorithms, like the Wiener Filter procedure. Afterward, we compared the neural network results with the results of applying the Wiener Filter with \textsc{NIFTy}. 

In the exact procedure of the Wiener Filter, the covariance matrices \textbf{S} and \textbf{N} are known \textit{a priori}.
The \textit{a posteriori} solution for the signal inference problem, $\hat{\bi{s}}$, is calculated through the Wiener Filter equation: 
\begin{equation}
    \hat{\bi{s}} = (\textbf{S}^{-1} + \textbf{R}^{\dag}\textbf{N}^{-1}\textbf{R})^{-1} (\textbf{R}^{\dag}\textbf{N}^{-1}\textbf{d}) = \textbf{D}j.
\label{eq:eq12}
\end{equation}

Then, the calculation of $\textbf{D} = (\textbf{S}^{-1} + \textbf{R}^{\dag}\textbf{N}^{-1}\textbf{R})^{-1}$ is performed numerically through the Conjugate Gradient algorithm. By default, conjugate gradient solves $D^{-1} \hat{\bi{s}} = j$, where $D^{-1}$ can be ill-conditioned (which is analyzed by the condition number of a non-singular and normal matrix). Depending on the power spectrum chosen, $\textbf{S}^{-1}$ can be responsible for the bad conditioning of D, therefore the election of a preconditioner (a non-singular matrix) is to solve the equivalent problem but better conditioned. A trivial choice for a preconditioner could be $\textbf{S}^{-1}$~\citep{chen_2005} to accelerate convergence.

\subsection{Power spectrum estimation}

Once we estimate the signal, $\hat{\bi{s}}$, we would like to compress this information into a summary statistic, like the power spectrum. If it is binned, the parameters defining the power spectrum can be specific band-powers $\bi{\Theta}$. This power spectrum, also interpreted as a covariance matrix, is employed as a prior in the Wiener Filter estimation described above. 
In this case, however, we can reconsider the Bayesian estimator that maximizes the conditional probability of the signal given the data, and marginalizes it over the signal $\bi{s}$. Once we have $P(\bi{\Theta}|\bi{d})$ we want to find the peak posterior solution and obtain an estimation of the power spectrum parameters $\Theta_\ell$, independent of the underlying field $\hat{\bi{s}}$.
In what follows, we adopt the approach outlined in~\cite{Seljak2017}.

We define a derivative matrix $\bi{\Pi}_\ell$ around some fiducial power spectrum $\bi{S}^{\rm fid}$ as
\begin{equation}
\left[\frac{\partial \bi{S} }{\partial \Theta_\ell}\right]_{\bi{S}^{\rm fid}}=\bi{\Pi}_\ell.
\label{pi}
\end{equation}
Note that for linear dependence of $\bi{S}$ on $\bi{\Theta}$ we can use 
\begin{equation}
\bi{\Pi}_\ell=\frac{\bi{S}_{\rm fid} }{\Theta_\ell},
\label{pilin}
\end{equation}
i.e. $\langle s_{k_\ell}s_{k_\ell}^* \rangle=\Theta_\ell\Pi_l(k_\ell)$.  
In this case, the derivative matrix takes us from $\Theta_\ell$, the power spectrum value representative over a bin, to $\bi{S}$ which is the power spectrum. 
In terms of this, we can write
\begin{equation}
\bi{S}=\bi{S}^{\rm fid}+\sum_\ell\Delta \Theta_\ell\bi{\Pi}_\ell.
\end{equation}

We assumed a flat prior on $\bi{\Theta}$, leading to the posterior $P(\bi{\Theta}|\bi{d})$ being proportional to the likelihood $L(\bi{\Theta}|\bi{d})$. We thus want to find the most probable $\bi{\Theta}$ given a set of measurements $\bi{d}$. To maximize the likelihood with respect to the bandpowers $\bi{\Theta}$, we expand the log-likelihood in terms of $\bi{\Theta}$ to  quadratic order around some fiducial values $\bi{\Theta}_{\rm fid}$,

\begin{equation}
\ln L(\bi{\Theta}_{\rm fid}+\Delta \bi{\Theta})=\ln L(\bi{\Theta}_{\rm fid})+\sum_\ell \left[\frac{\partial \ln L(\bi{\Theta})}{\partial \Theta_\ell} \right]_{\bi{\Theta}_{\rm fid}}\Delta \Theta_\ell+ \frac{1}{2} \sum_{\ell \ell'}\left[\frac{\partial^2 \ln L(\bi{\Theta})}{\partial \Theta_\ell \partial \Theta_{\ell'}}\right]_{\bi{\Theta}_{\rm fid}}\Delta \Theta_\ell \Delta \Theta_{\ell'}.
\label{llik}
\end{equation}
Note that the last term of equation \ref{llik} defines the curvature matrix as the second 
derivatives of log-likelihood with respect to the parameters. 
Following ~\cite{Seljak2017}, we define
\begin{equation}
E_\ell(\bi{S}_{\rm fid},\hat{\bi{s}})=\frac{1}{2}\hat{\bi{s}}^{\dag}\bi{S}_{\rm fid}^{-1}\bi{\Pi}_\ell\bi{S}_{\rm fid}^{-1}\hat{\bi{s}} = \frac{1}{2}\sum_{k_\ell}\frac{\hat{s}_{k_\ell}^2 }{\Theta_{\rm{fid},\ell} S_{{\rm fid},k_\ell}},
\label{el}
\end{equation}
where the sum over $k_\ell$ accounts for all the modes that contribute to the bandpower $\Theta_\ell$. In the last equality, we use the diagonal property of the projection operators and of the fiducial power spectrum. With this expression, the first derivative of the likelihood becomes
\begin{equation}
\frac{\partial \ln L(\bi{\Theta}) }{\partial \Theta_\ell}=E_\ell - b_\ell .
\label{like_deriv}
\end{equation}
Then, the maximum likelihood solution for $\bi{\hat{\Theta}}$ is obtained setting the equation above equal to zero leading to the noise bias term equal to:
\begin{equation}
b_\ell= E_\ell(\bi{\Theta}_{\rm fid},\hat{\bi{s}}_{s+n}),
\label{blmv}
\end{equation}
where $\hat{\bi{s}}_{s+n}$ is the Wiener-filtered map obtained from data that contains signal and noise ($s+n$), where the signal map is a realization of the fiducial power spectrum.

Instead of solving the equation directly, the Newton's method is employed,
in which the solution of the quadratic log-likelihood at the peak can be found by setting the derivative of \ref{llik} with respect to $\Delta \bi{\Theta}$ equal to zero, and using the equation \ref{like_deriv}, leading to: 
\begin{equation}
    (\bi{F}\Delta\bi{\Theta})_{\ell} = E_\ell - b_\ell,
    \label{corr}
\end{equation}
where the signal map appearing in $E_\ell$ is a realization of the true power spectrum, and  $F$ is the Fisher matrix defined as the ensemble average of the curvature around the maximum, 
\begin{equation}
F_{\ell\ell'}=-\left\langle \frac{\partial^2 \ln L(\bi{\Theta})}{ \partial \Theta_\ell \partial \Theta_{\ell'}} \right\rangle .
\label{fish1}
\end{equation}

In the case of  gaussian-distributed modes, the inverse of the Fisher matrix $\bi{F^{-1}}$ can be interpreted as the covariance matrix of the estimated parameters $\bi{\hat{\Theta}}$:
\begin{equation}
    \bi{F^{-1}} = \langle\Delta\bi{\hat{\Theta}}\Delta\bi{\hat{\Theta}}^{\dagger}\rangle - \langle\Delta\bi{\hat{\Theta}}\rangle \langle\Delta\bi{\hat{\Theta}}^{\dagger}\rangle. 
\end{equation}

The reconstruction $\bi{\hat{s}}$ contains noise contribution. Squaring it for the calculation of the raw power spectrum $E_{\ell}$ will require the subtraction of a noise bias term to obtain an unbiased estimator, as stated in equation \ref{corr}. Then, we obtain the bandpower estimates convolved with the Fisher matrix $\bi{F}\Delta\bi{\hat{\Theta}}$, where the Fisher matrix describes both the covariance matrix and the bandpower mixing, since we are not considering a complete coverage in the estimation of the signal.

The bias term and the Fisher matrix are calculated with simulations as
we will describe in section~\ref{sec:results_hom}. Lastly, the estimation of the power spectrum will be $\bi{\Theta}_{fid} + \Delta\bi{\hat{\Theta}}$.

\section{Neural Network for the Wiener Filter}
\label{sec:CNN}

Machine learning (ML) methods can be broadly categorized into two types: supervised and unsupervised learning. In supervised learning, models are trained with examples consisting of features and target values (i.e., the actual values that the models aim to predict). The model is constructed using a prediction function that combines these features to generate predictions, which are then compared to the targets using a loss function. Learning involves finding optimal weights that associate features with labels, and minimizing the loss function to enable accurate predictions. In contrast, unsupervised methods are optimized using tailored loss functions that are mathematically guaranteed to be minimized by the exact solution, without the need for labeled data during training.

\subsection{Model training, inference, and predictions}

Machine learning models comprise two primary phases: training and inference. During training, the algorithm iteratively minimizes the cost function, often employing techniques like gradient descent, to learn the model's weights or parameters. This cost function quantifies the difference between the model's predictions and the training examples' targets.

In the inference phase, the trained model is evaluated by making predictions on new, unseen data. Generalization refers to the model's ability to adapt to new data not encountered during training. Overfitting, which happens when the model fits the training data perfectly but performs poorly on new data, is addressed by splitting the data into training and validation sets. The training set is used for model training, while the validation set assesses performance after each training epoch using the acquired weights.

Following the training phase, the model makes predictions on a distinct dataset called the test set, consisting of new examples. The selected model is determined by its optimal performance on the validation set, and its prediction results are validated using the test set. Hyperparameters, which are parameters influencing the model's performance, are carefully chosen by analyzing the model's behavior on the validation set, typically using metrics such as validation loss or precision.

\subsection{Convolutional neural networks}

Convolutional Neural Networks (CNNs) are a class of neural networks known for their effectiveness in image analysis and recognition tasks. They employ complex structures and nonlinear combinations of features to process visual data. Activation functions can be added to introduce nonlinearity~\citep{Chollet2017} into the network. 
CNNs consist of two-dimensional layers with connections between neurons organized in receptive fields. These networks use kernels or filters to detect specific image characteristics, which generate feature maps. As a result, each layer of the neural network becomes three-dimensional.

The CNN architecture enables the creation of a convolutional autoencoder, consisting of an encoder and a decoder. The encoder transforms inputs into an efficient latent space representation, while the decoder maps this representation back to the expected output. One example of such an architecture is UNet~\citep{UNet},  which connects encoders through skip connections to facilitate the transfer of features. During the encoder phase, layer dimensions are successively reduced by a factor of 2 using a stride of 2 in the convolutional layers. In contrast, the decoder increases layer dimensions, ultimately producing an image of the same size as the inputs. For more details about the basic operations behind the network see Appendix \ref{apx2}.

WienerNet~\citep{Munchmeyer2019},  an architecture inspired by UNet, is a neural network designed to simulate the Wiener Filter. It takes noisy CMB images as inputs and produces reconstructed maps of the original signal. The architecture consists of two channels: the first channel analyzes the noisy CMB image using convolutional layers, while the second channel processes the mask, identifying regions with infinite noise, using nonlinear convolutional layers with a Rectified Linear Unit (ReLU) activation function~\citep{Goodfellow2016}. The outputs of the mask channel are then multiplied with the outputs of the map channel.


The encoder part of the architecture applies convolutional layers to the image, using multiple filters to extract key features and generate a latent representation of the image. Subsequently, the decoder part upscales the image, returning a map with the same dimensions as the original input.

The Wiener Filter matrix, $M$, in this context, represents the weights learned during the training of the neural network. The exact WF of the CMB map is not included as a target in the training set due to its computationally expensive nature, instead, the loss function used in WienerNet
is motivated by the posterior likelihood of the true CMB sky given noisy data realization and is defined as:
\begin{equation}
J(d,y) = \frac{1}{2} (y-d)^{T}N^{-1}(y-d) + \frac{1}{2}y^{T}S^{-1}y,
\end{equation}
where $y$ represents the predicted output (which is an estimate of the reconstructed signal, i.e., the Wiener filtered map), $d$ is the noisy data, and $N$ and $S$ are covariance matrices associated with noise and signal, respectively. 

By minimizing the loss function $J$, which is equivalent to maximizing the posterior likelihood, the learned weights in the matrix $M$ converge to the Wiener filter matrix as defined in Equation \ref{eq:eq10} (with $R=I$). This ensures that, after training, the model effectively simulates the behavior of the Wiener filter when making predictions on new data. For further details on the minimization of $J$, refer to the work of~\cite{Munchmeyer2019}. The implementation of $J$ is: 

\begin{equation}
    J = \sum_{i}^{N_{pix}} \frac{(T_{i}^{NN}-T_{i}^{obs})^{2}}{N_{i}} + \sum_{\ell} \frac{T_{\ell}^{NN}T_{\ell}^{NN*}}{C_{\ell}}, 
\end{equation}
where $T^{obs}$ is the input map itself, $T^{NN}$ is the output map of the neural network, and $N_{i}$ is the pixel noise variance of the experiment. The first term is evaluated in pixel space while the second term is evaluated in Fourier space. 

\begin{figure}
\centering
\includegraphics[width=1.\columnwidth]{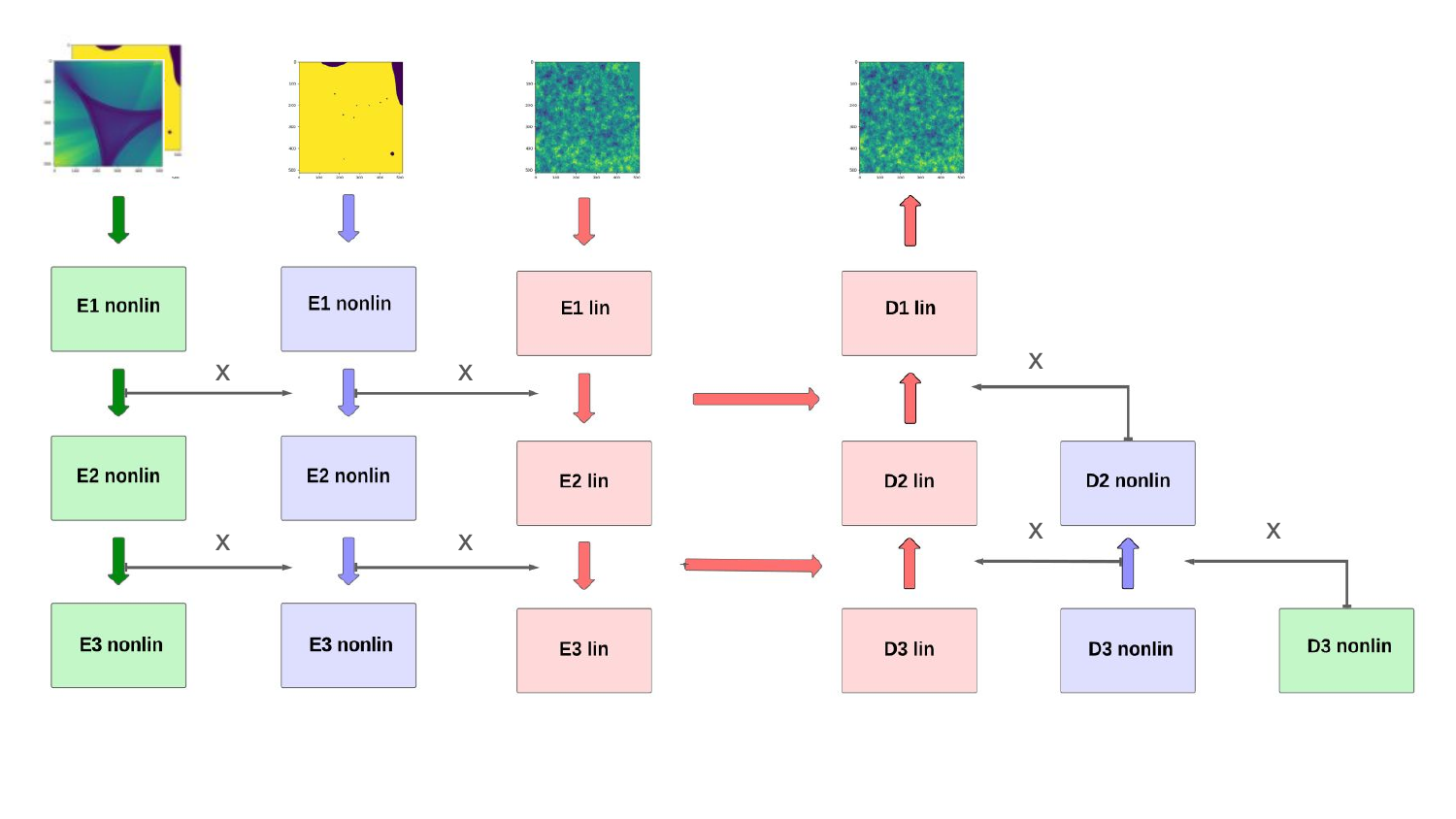}
\caption{Scheme of the neural network architecture: only three encoders are presented for visualization purposes.  The black solid line represents the multiplication between channels, and the red arrow connecting the encoder part to the decoder part represents the skip connections in the linear channel. These skip connections are present also in the other channels although not being drawn here. 
The third channel is two-dimensional, containing both the variance map and the mask.
}
\label{esquema}
\end{figure}

Besides migrating WienerNet from \textsc{Tensorflow1} to \textsc{Tensorflow2} for the homogeneous noise case, we have also extended the neural network architecture for addressing the inhomogeneous noise problem. This involved adding a third channel for the inhomogeneous map variance, in addition to the channel for the mask already present in WienerNet\footnote{We have also explored an alternative architecture, with only one extra channel, in which the information of the mask was included in the variance map as regions of infinite noise. We found that the training performance was worse in this case, so we did not implement it in the final neural network.}. 
The additional channel operates similarly to the mask channel, multiplying the results of the linear map channel, and implementing skip connections in every channel. Figure \ref{esquema} illustrates a simplified representation of this new channel within the network architecture.
It was also noted that treating the third channel as two-dimensional, including both the variance map and the mask, significantly improves the efficiency of the neural network training convergence. Hence, this two-dimensional approach for the third channel is adopted in this paper.
Refer to Appendix \ref{apx2} for more details about the neural network structure.

\section{Datasets}
\label{sec:datasets}
In order to train neural networks, it is essential to generate data maps that accurately represent the properties of both signal and noise. In this study, we have considered different cases: homogeneous noise, and inhomogeneous noise.

\subsection{Maps}

In our study, we adopt the flat sky approximation since we are working with small sky regions. The angular resolution of the map is determined by the size of the map and the number of pixels used for its partition.

We analyzed maps of different sizes ($56 \times 56$, $128 \times 128$, $256 \times 256$, $512 \times 512$) to assess the computational time for WienerNet predictions in comparison to \textsc{NIFTy}'s Wiener Filter application. Increasing the map size implies a modification of the neural network's architecture adding more encoders and decoders in order to obtain the desired output shape.
The duration of the training of the neural network is significantly affected by the map size, as shown in~\cite{Belen_BAAA2023}. In Appendix~\ref{apx1}, we provide a time-scaling analysis for the training and execution of the neural network, comparing these times with the exact Wiener Filter calculation using \textsc{NIFTy}.

For maps sized at $5^\circ \times 5^\circ$ and $128 \times 128$ pixels, we trained various WienerNet models with different hyperparameters, using a \textit{fiducial} angular power spectrum for the generation of the training maps.
We begin by interpolating the values of the angular power spectrum $C_\ell$ onto a 2D power spectrum grid for the flat sky.
 The noise in this case is homogeneous, with uniform variance across all map pixels. 
The trained neural network's weight matrix represents the Wiener Filter matrix (eq~\ref{eq:eq10}).
 These models are then used to estimate the power spectrum of maps with a different \textit{true} angular power spectrum, which were not part of the training data.

Additionally, we trained the neural network with maps sized at $20^\circ \times 20^\circ$ and $512 \times 512$ pixels, incorporating inhomogeneous noise. We simulated maps with various inhomogeneous noise models, creating distinct neural network models for each of them.
We evaluated each model's performance in simulating the Wiener Filter for the corresponding noise model and proceeded with power spectrum estimation, following a similar procedure as in the homogeneous case. 
The \textit{fiducial} power spectrum was obtained using \textsc{CAMB}~\cite{CAMB}, with the cosmological parameters from the best fit of Planck data (2014), obtained from Table 5 in~\cite{planck2013}. The  \textit{true} power spectrum was obtained similarly but with the cosmological parameters changed by 4$\sigma$ from their fiducial values.
Each neural network used in this work was trained with the Apollo GPU nodes of the Institute for Advanced Studies, in which every node contains 8 GPUs NVIDIA A100.

\subsection{Mask}

\begin{figure}[ht!]
\centering
\includegraphics[width=1.\columnwidth]{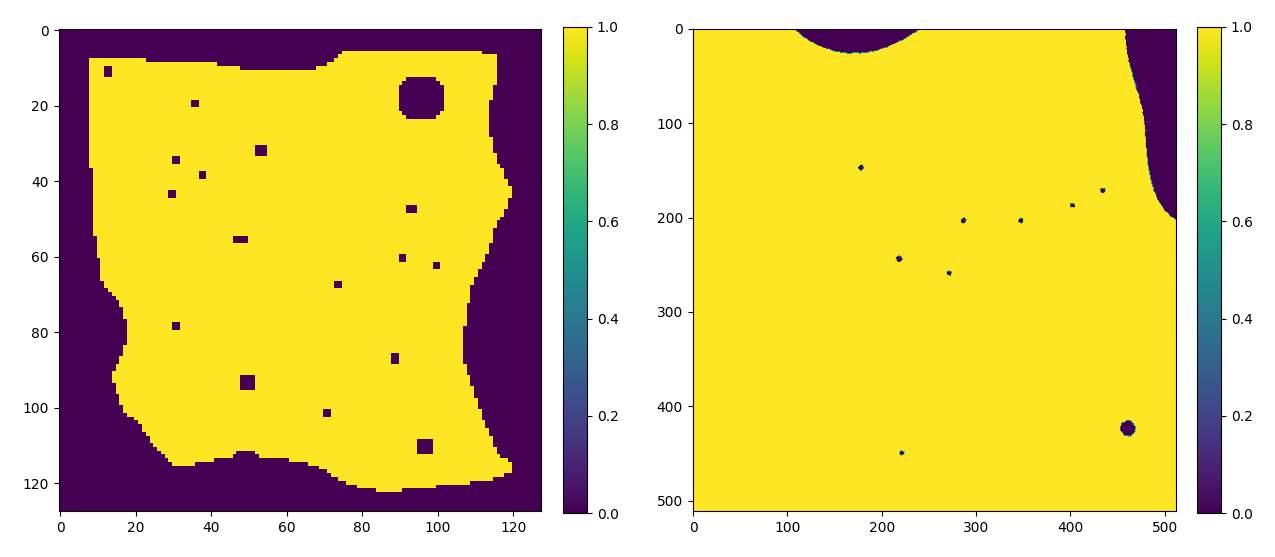}
\caption{Left panel: mask applied to the noisy data for the homogeneous problem. Right panel: mask applied to the noisy data for the inhomogeneous problem.}
\label{mask}
\end{figure}

In Figure \ref{mask}, we show the mask applied to the noisy maps for homogeneous and inhomogeneous cases. For the homogeneous case (left panel) we used the same mask as in~\citep{Munchmeyer2019}, while for the inhomogeneous case (right panel) we extracted a mask sample from Planck maps~\citep{Planck2020} as an example, containing point sources and extended regions. Our algorithm allows the use of more complicated masks, by sending them as an argument in the data simulations code for the inhomogeneous case.

\subsection{Noise model}

To evaluate the efficiency of WienerNet  
with respect
to the CG (conjugate gradient) method and explore how results scale with the number of pixels and 
for different noise models, we simulate homogeneous noise models at three different noise levels for each of the specified map sizes. For detailed information about the simulation of the homogeneous model for efficiency study purpose refer to Appendix~\ref{apx1}.

\begin{figure}
\centering
\includegraphics[width=1.\columnwidth]{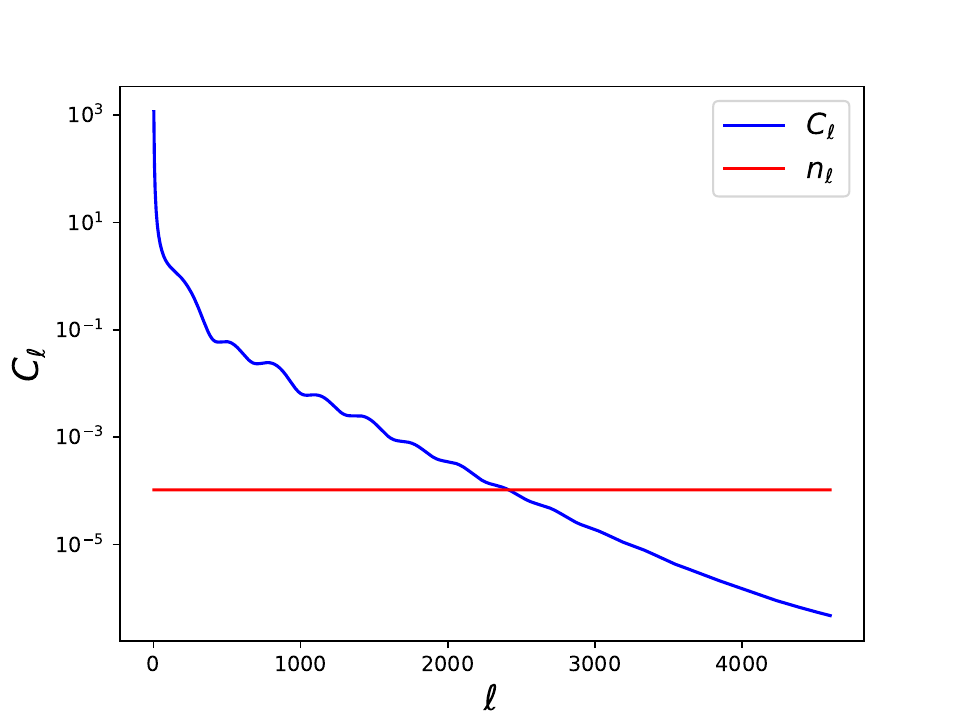}
\caption{Power spectrum of the signal and power spectrum of a gaussian white noise. A noise level of $35 \mu$K-arcmin corresponds to a noise power spectrum at the level of  $n_\ell = 0.0001036$.}
\label{homogeneo}
\end{figure}

To assess the performance of WienerNet compared to the CG method, utilizing a version of \textsc{Tensorflow} different from the one in \citep{Munchmeyer2019}, and to perform power spectrum estimation, we focused on maps with dimensions of $128 \times 128$ pixels and a homogeneous noise level of $35 \mu K-arcmin$, as presented in Figure~\ref{homogeneo}. The scale at which noise dominates the signal and the spectrum transitions to noise-dominated occurs at $\ell = 2415 $.

\begin{figure}
\centering
\includegraphics[width=0.95\columnwidth]{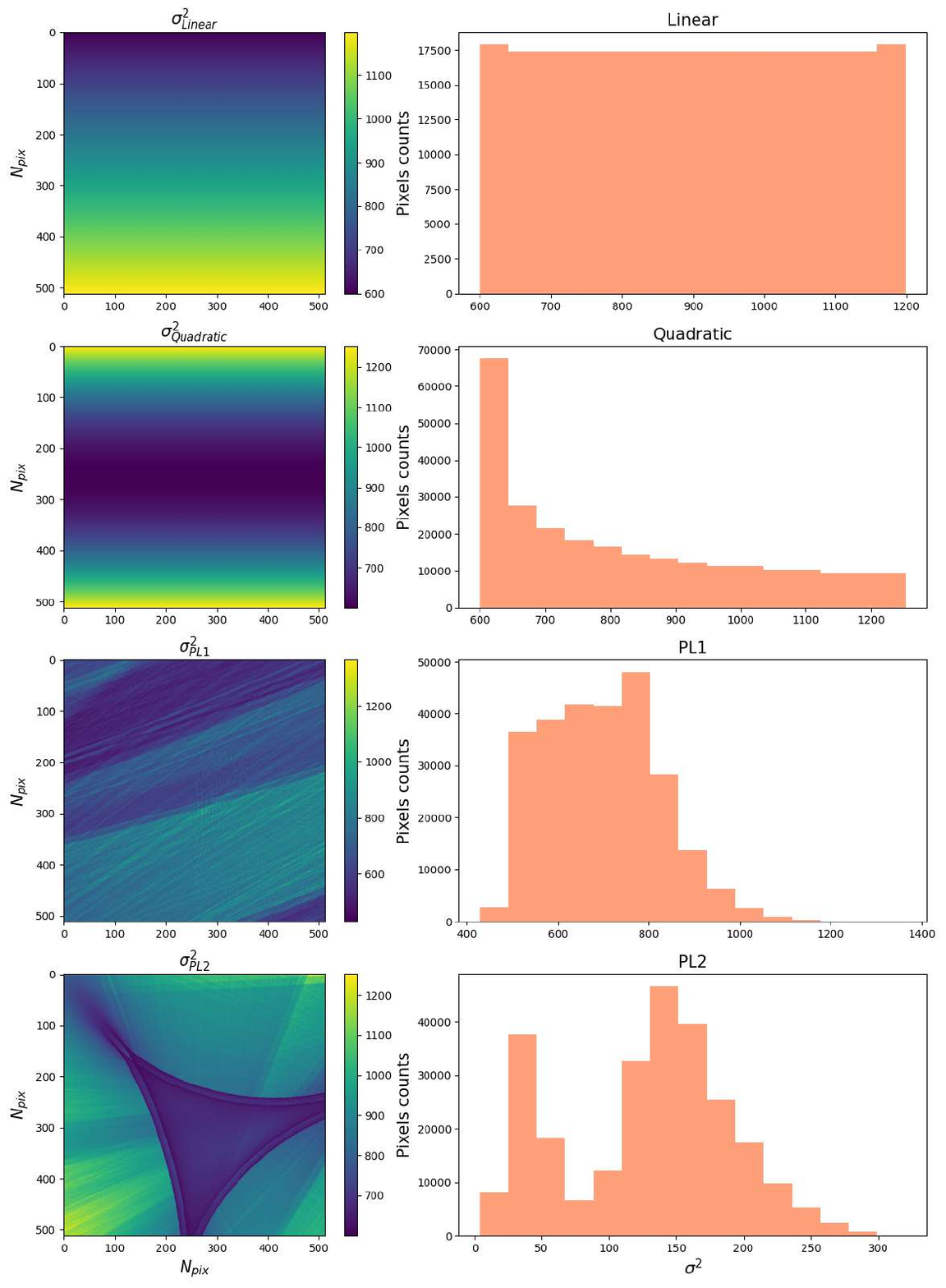}
\caption{Left panel: Variance maps generated from linear and quadratic functions, and the variance samples extracted from Planck. Right panel: Histogram of each variance map showing the distribution of pixel counts
across different noise variances.}
\label{inho_hist}
\end{figure}

For the inhomogeneous case, we introduced gaussian noise with varying variance across the map. We generated variance maps from linear and quadratic functions and we also considered realistic noise samples obtained from Planck noise maps~\citep{Planck2020}, as illustrated in the left panel of Figure~\ref{inho_hist}, to assess the performance of the neural network with more complex variance maps. The pixel counts for
 different noise variance values,  for each variance map, is shown in the histograms on the right panel of Figure \ref{inho_hist}. From the histograms of the Planck variance values, it is evident that the second variance map (Planck 2) exhibits a low average noise level but features a more complex structure and a broader range of variance values across pixels. 

\begin{figure}
\centering
\includegraphics[width=1.\columnwidth]{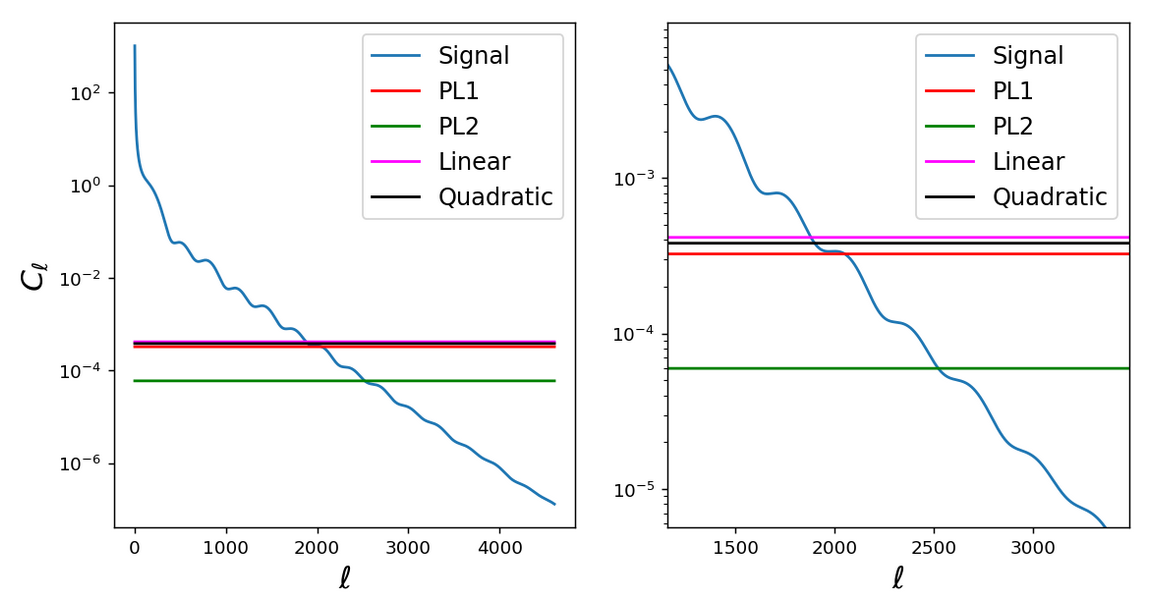}
\caption{Left panel: power spectrum of the signal for $512 \times 512$ maps and average noise for four inhomogeneous variance maps. Right panel: detailed into the transition angular scales
from a signal-dominated regime to a noise-dominated regime. From this Figure, it is clear the scales where the power spectrum of the average noise cut the signal. 
}
\label{noise_espectros}
\end{figure}

In Figure ~\ref{noise_espectros}, we present the power spectrum of the signal and the average noise levels for the four inhomogeneous noise scenarios presented in Figure \ref{inho_hist}. A closer look reveals that the noise levels in each case cut the signal at different scales. Specifically, for the linear and quadratic variance map, the noise intersects the signal at scales $\ell = 1879$ and $\ell = 1900$, respectively, while for the Planck variance maps, it occurs at scales $\ell = 2050$ for Planck 1 and $\ell = 2521$ for Planck 2.

\begin{figure}
\centering
\includegraphics[width=1.\columnwidth]{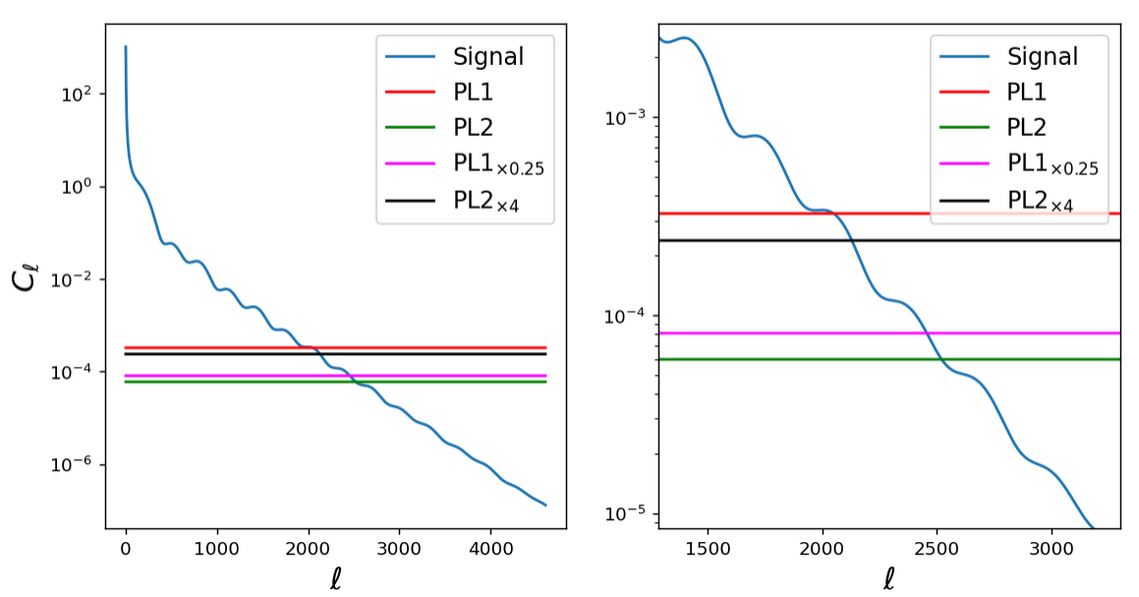}
\caption{Left panel: power spectrum of the signal for $512 \times 512$ maps and average noise of Planck 1 and 2, also scaled by a factor 4. Right panel: zoom in on the transition from a signal-dominated regime to a noise-dominated regime. It is clear from the plot the scales where the power spectrum of the average noise cut the signal.}
\label{espectros_escaled}
\end{figure}

The number of modes with a high signal-to-noise ratio varies among cases due to the average noise model affecting the signal at different scales. Surprisingly, Planck 2, despite being the most inhomogeneous noise case, displays a higher number of modes with a high signal-to-noise ratio. To explore the impact of noise level and the complexity in the distribution of variable variances, we scaled the Planck variance maps. Multiplying Planck 2 by a factor of 4 shifted the scale of intersection to $\ell=2130$, while dividing Planck 1 by the same factor shifted the scale to $\ell=2457$, as shown in the power spectrum of these maps in Figure \ref{espectros_escaled}. In Table \ref{tab:table1} we present the notation that we will use from now on to specify each noise variance map.

This analysis helps us understand whether the neural network's convergence is more influenced by the average noise level or by the complexity in the distribution of variable variances in realistic inhomogeneous problems. We describe our findings in the following sections.
\begin{table}[ht!]
\centering
\label{tab:table1}
\begin{tabular}{|c |c |} 
\hline
Notation & Inhomogeneous variance map \\ [0.5ex] 
\hline
Linear & Variance map generated from a linear function \\ 
Quadratic & Variance map generated from a quadratic function \\
PL1 & Planck variance map 1 \\
PL1$_{\times0.25}$ & Planck variance map 1 divided by 4 \\
PL2 & Planck variance map 2 \\
PL2$_{\times4}$ & Planck variance map 2 multiplied by 4 \\ [1ex] 
\hline
\end{tabular}
\caption{Notation to specify each noise variance map used to generate gaussian inhomogeneous noise.}
\end{table}

\section{Results: Homogeneous case}
\label{sec:results_hom}

\subsection{Comparison with \textsc{NIFTy} results}

For the homogeneous noise case, we trained our own version of WienerNet in \textsc{Tensorflow~2} and \textsc{Python~3} for maps with $128 \times 128$ pixels. We trained several models with different hyperparameters and sizes of the training set in order to study how well the performance of WienerNet on simulating the WF regarding the optimization of the neural network is. In Table \ref{tab:table3} it is presented two trained models with different sets of hyperparameters.
\textcolor{black}{These values were selected to ensure that the validation loss function is minimized.} 

\begin{table}[ht!]
\centering
\begin{tabular}{|c |c |c |} 
\hline
 { } & Model 1 &  Model 2 \\ 
\hline
\# training maps & 4000 &  20000 \\
\# validation maps & 1000 & 1000  \\
learning rate & 0.001 &  0.0001 \\
Batch size & 8 &   8 \\ 
Optimizer & Adam V1 & Adam V2 \\
Initializer & Glorot uniform &  Random uniform \\
 \# filters & 10 & 10 \\ 
\hline
\end{tabular}
\caption{Hyperparameters and dataset set size (training and validation) used for two different models. The last line indicates the number of filters used per layer.}
\label{tab:table3}
\end{table}

We started with Model 1 using the Adam~\citep{2014arXiv1412.6980K} version compatible with TF1 because the initial version of WienerNet presented in~\citep{Munchmeyer2019} was developed in \textsc{Tensorflow~1.11}, in which we were able to reproduce the results. We changed the initialization of the weights from Glorot uniform~\cite{article} to Random uniform and also the size of the training set, not reaching a better performance in the neural network model. Then we decided to use the optimizer Adam version 2, modifying the number of maps in the training set and the learning rate, finding that the best performance of the neural network is obtained with the hyperparameters presented in Model 2.

The agreement in the WF simulated with WienerNet compared to the exact WF method can be established using the cross-correlation coefficient $r_{\ell}$ as a function of the multipole $\ell$: 
\begin{equation}
    r_{\ell} = \frac{\langle a_{CNN}(\ell) a^*_{WF}(\ell)\rangle}{\sqrt{\langle a_{CNN}(\ell) a^*_{CNN}(\ell)\rangle \langle a_{WF}(\ell) a^*_{WF}(\ell)\rangle}},
\end{equation}
where $a_{CNN}$ and $a_{WF}$ are the discrete Fourier coefficients of the output map of the neural network and the exact WF respectively, and $r_{\ell}$ is averaged over the test set (100 maps).
~ 
\begin{figure}
\centering
\includegraphics[width=1\columnwidth]{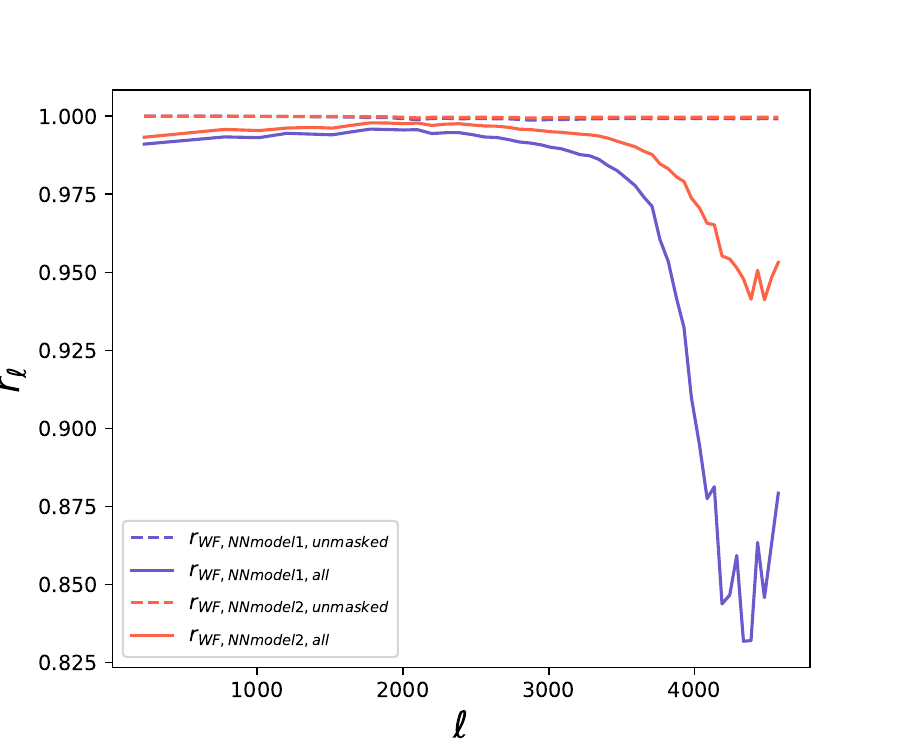}
\caption{
Cross-correlation coefficient using the neural network trained with the hyperparameters specified in Model 1  (blue solid line) and with the hyperparameters specified in Model 2 (red solid line).  We note a clear decorrelation near  $\ell = 3500$. The dotted lines represent the cross-correlation coefficient in the unmasked region, for both models, which is equal to 1 in all scales.
}
\label{fig:correlaciones}
\end{figure}

Figure \ref{fig:correlaciones} shows the cross-correlation coefficient between WienerNet and the exact WF using the set of hyperparameters specified above. In both cases, a decorrelation is observed at the scale $\ell = 3500$, where the noise level is several orders of magnitude greater than the signal. However, for a neural network trained with a larger training set and a different version of the Adam optimizer (Model 2), the decorrelation is less pronounced at those scales. Nevertheless, the observed decorrelation of the neural network compared to the exact Wiener Filter comes from the masked region, as represented by the dotted lines in Figure \ref{fig:correlaciones}. In this case, the cross-correlation coefficient was calculated in the unmasked region and remains equal to 1 across all scales. 

\begin{figure}
\centering
\includegraphics[width=1\columnwidth]{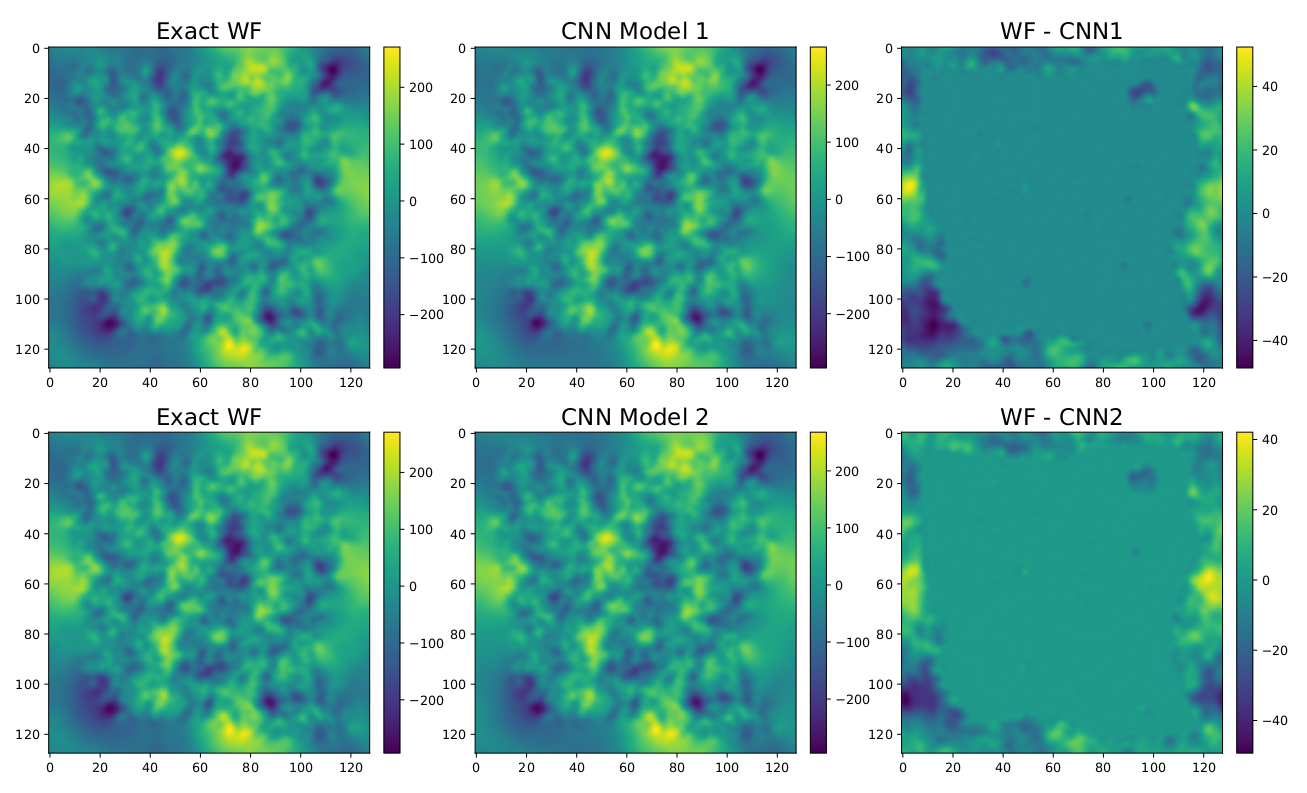}
\caption{On the left is the output map of the exact WF calculated with \textsc{NIFTy}, in the center the output map calculated with WienerNet (Model 1 in the top panel, Model 2 in the bottom panel), and on the right is the difference between the two.
}
\label{fig:residuo1}
\end{figure}

In Figure \ref{fig:residuo1}, the first and second panels show the filtered map with the exact WF and the neural network, respectively, for a specific example of a signal map. \textcolor{black}{In both methods, the reconstruction of large-scale modes in the masked region, particularly at the border, is clear, while the reconstruction of small-scale modes is comparatively less noticeable. This produces a ''blurring effect'' on the reconstructed map.} The third panel shows that the pixel difference between the filtered maps produced by WienerNet and the exact WF is more pronounced in the masked region. This observation holds true for both WienerNet optimization models (Model 1 and Model 2).

\begin{figure}
\centering
\includegraphics[width=1\columnwidth]{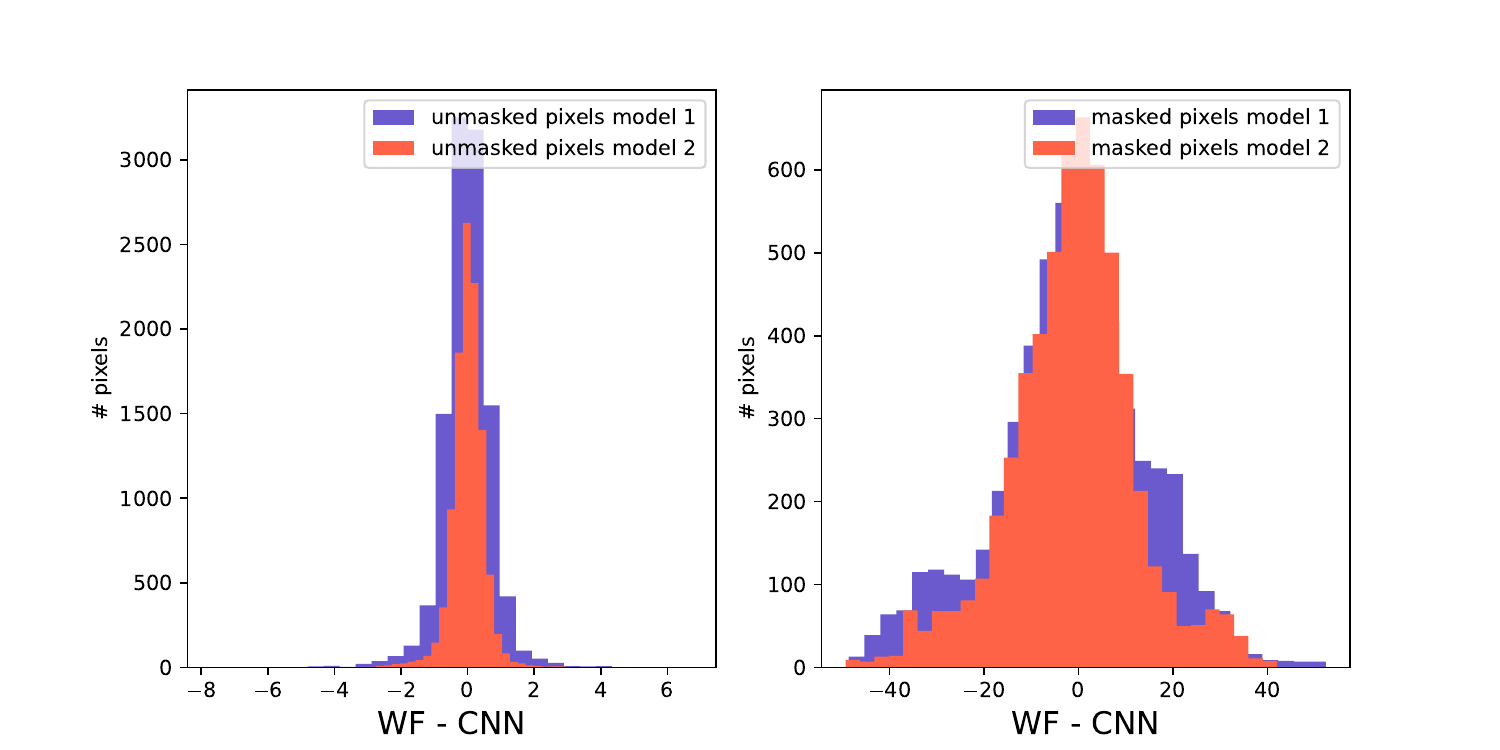}
\caption{Left panel: Histogram of the map difference between exact WF and WienerNet in unmasked region with Model 1 and 2. Right panel: Histogram of the map difference between exact WF and WienerNet in masked region with Model 1 and 2  with greater values than the unmasked pixels}.
\label{fig:histograma1}
\end{figure}

For that specific map example, the difference between the exact WF and WienerNet with Model 1 has a mean equal to $-0.58$ and a standard deviation equal to $9.54$. At the same time, the difference between the exact WF and WienerNet with Model 2 has a mean equal to $-0.37$ and a standard deviation equal to $7.88$. The histograms illustrating these map differences are shown in Figure \ref{fig:histograma1}. 

It is noteworthy that the unmasked pixels exhibit much smaller differences compared to the masked pixels, from the color bar in the maps of Figure \ref{fig:residuo1} and the magnitudes of the histogram \ref{fig:histograma1} it is evident that the difference is the order of $1\%$ and $10\%$ for the unmasked and masked pixels respectively. 

These results are aligned with the cross-correlation coefficient results in Figure \ref{fig:correlaciones}. The coefficient remains approximately equal to 1 across all scales in the unmasked region but shows decorrelation at small scales (high values of $\ell$) when we consider the masked region.
~

The cross-correlation coefficient assesses the correlation between the exact WF map computed using \textsc{NIFTy}, which employs a conjugate gradient method with a preconditioner, and the filtered map generated by the trained neural network WienerNet. The WienerNet Model 2 exhibits less decorrelation in the masked region at smaller scales. Consequently, we have selected this model as the neural network model to use in the power spectrum estimation procedure. However, from the histograms and figures presented earlier, it is evident that, even though the decorrelation of WienerNet Model 1 is higher at smaller scales within the masked region, the reconstruction of the signal with both models is similar, making both models feasible for use.

In the next section, we will explore whether the performance of the WienerNet model in \textsc{TensorFlow~2} has any impact on the estimation of the power spectrum.

\subsection{Implementation of the power spectrum}
\label{sec:powerspectrum_hom}

For the estimation of the power spectrum, it is necessary to assess the noise bias and the Fisher matrix, presented in section~\ref{sec:Background}. This evaluation takes place after the estimation of the underlying field $\hat{s}$, for which the selection of a \textit{fiducial} angular power spectrum is required. Subsequently, we proceed to estimate the unknown power spectrum referred to as the \textit{true}, which is unbiased and incorporates the correction window function describing correlation between the bandpowers $\Theta_{\ell}$. 

We computed the noise bias and Fisher matrix through simulations of the fiducial power spectrum. We began generating a gaussian random realization of the signal in Fourier space, denoted as $\bi{s}_{s}$, as the fiducial model:
\begin{equation}
    \langle |\bi{s}_{s}|^2 \rangle = \bi{S}_{fid}.
\end{equation}

We generated a random noise realization using the noise level for the homogeneous case specified in section \ref{sec:datasets} and built a data vector:
\begin{equation}
    \bi{d}_{s+n} = \bi{d}_{s} + \bi{d}_{n}.
\end{equation}

We performed the WF estimation of the data simulation using the models of WienerNet described above, resulting in $\bi{\hat{s}}_{s+n}$. In order to obtain an unbiased estimator of the power spectrum we used the expression of the noise bias \ref{blmv} averaged over several realizations.  

For the Fisher matrix, we introduced a small perturbation to a Gaussian realization of the \textit{fiducial} power spectrum, injecting a minor amount of power in phase to the Fourier modes at a single bandpower $\ell'$, 
\begin{equation}
    \bi{s}_{\ell',s} = \bi{s}_{s} + \Delta \bi{s}_{\ell'}.
\end{equation}

Subsequently, the equation of the Fisher matrix averaged over several simulations, is derived from~\citep{Seljak2017}: 
\begin{equation}
    F_{\ell \ell'}\Delta \Theta_{\ell'} = E_{\ell}(\Theta_{fid}, \bi{\hat{s}}_{\ell',s+n})-E_{\ell}(\Theta_{fid}, \bi{\hat{s}}_{s+n}).
\label{eq:fisher}
\end{equation}

The Fisher matrix is interpreted here as a response of a bandpower $\ell$ to another bandpower at $\ell'$.  

\begin{figure}
\centering
\includegraphics[width=1\columnwidth]{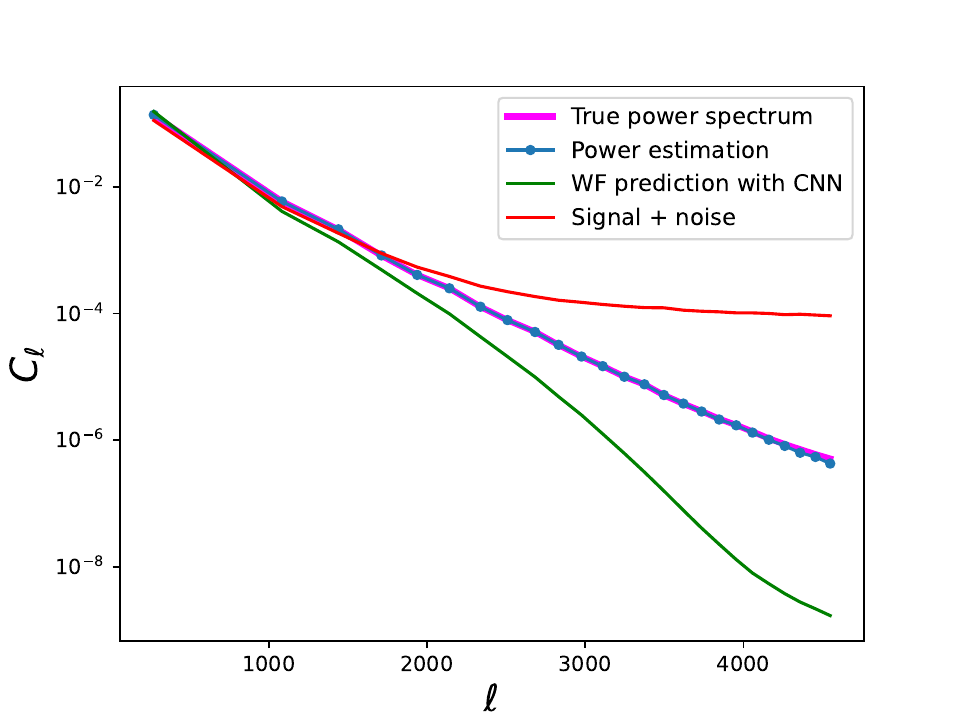}
\caption{
The magenta solid line is the \textit{true} power spectrum that we aim to estimate, and the red solid line is the power spectrum of the data. The green solid line is the power spectrum of the WF-filtered map with the CNN, where it is evident that it tends toward zero for high noise scales. The blue dotted line represents the estimation of the power spectrum after the WF estimation of the map and the procedure described for recovering the true signal.
}
\label{fig:est_homogeneo}
\end{figure}

After evaluating the noise bias and the Fisher matrix through Monte Carlo simulation approach, we generated map simulations of the \textit{true} angular power spectrum and a data sample, then we estimated the bandpower correction with the equation \ref{corr}. In Figure \ref{fig:est_homogeneo} the dotted line represents the average estimation of the \textit{true} power spectrum on 100 maps through the procedure described. The red solid line represents the power spectrum of the data sample (signal + noise) and the green solid line represents the WF power spectrum after estimation on the maps using the CNN. It is evident that for modes beyond the noise level, the WF estimation tends toward zero, in contrast to the modes in the signal-dominated regime, which closely align with the original signal.
~

\begin{figure}
\centering
\includegraphics[width=1\columnwidth]{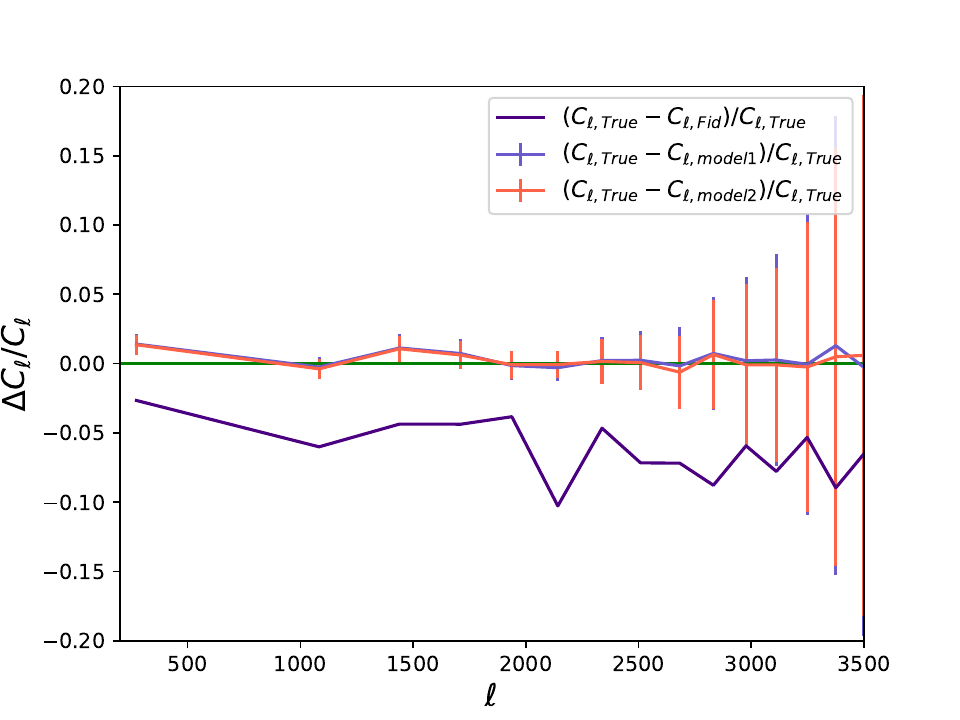}
\caption{The relative difference between the \textit{true} power spectrum and the power spectrum estimation using WienerNet models 1 and 2 is around zero and the relative error bars become larger for scales beyond $\ell = 3000$. Also, it presents the relative difference between the \textit{true} power spectrum and the proposed \textit{fiducial} power spectrum.}
\label{fig:relativ_homogeneo}
\end{figure}

To compare the ability of the two WienerNet models presented in the above section to estimate the \textit{true} power spectrum after this procedure, we analyze the relative difference between the power spectrum estimation and the \textit{true} power spectrum, as shown in Figure \ref{fig:relativ_homogeneo}. \textcolor{black}{The purple line represents the relative difference between the \textit{true} and \textit{fiducial} power spectra.
These power spectra are different by construction, in this case, by  5$\%$. The fiducial power spectrum serves as a reference, and as a starting point from which we can compute the unknown true power spectrum through the estimation procedure applied to a single map. } We see that the estimation of the true power spectrum from a proposed \textit{fiducial} power spectrum is unbiased for both models, and the relative error bars become larger when the level of noise is too high compared to the signal, specifically for modes beyond $\ell = 3000$. 
~

\begin{figure}
\centering
\includegraphics[width=1\columnwidth]{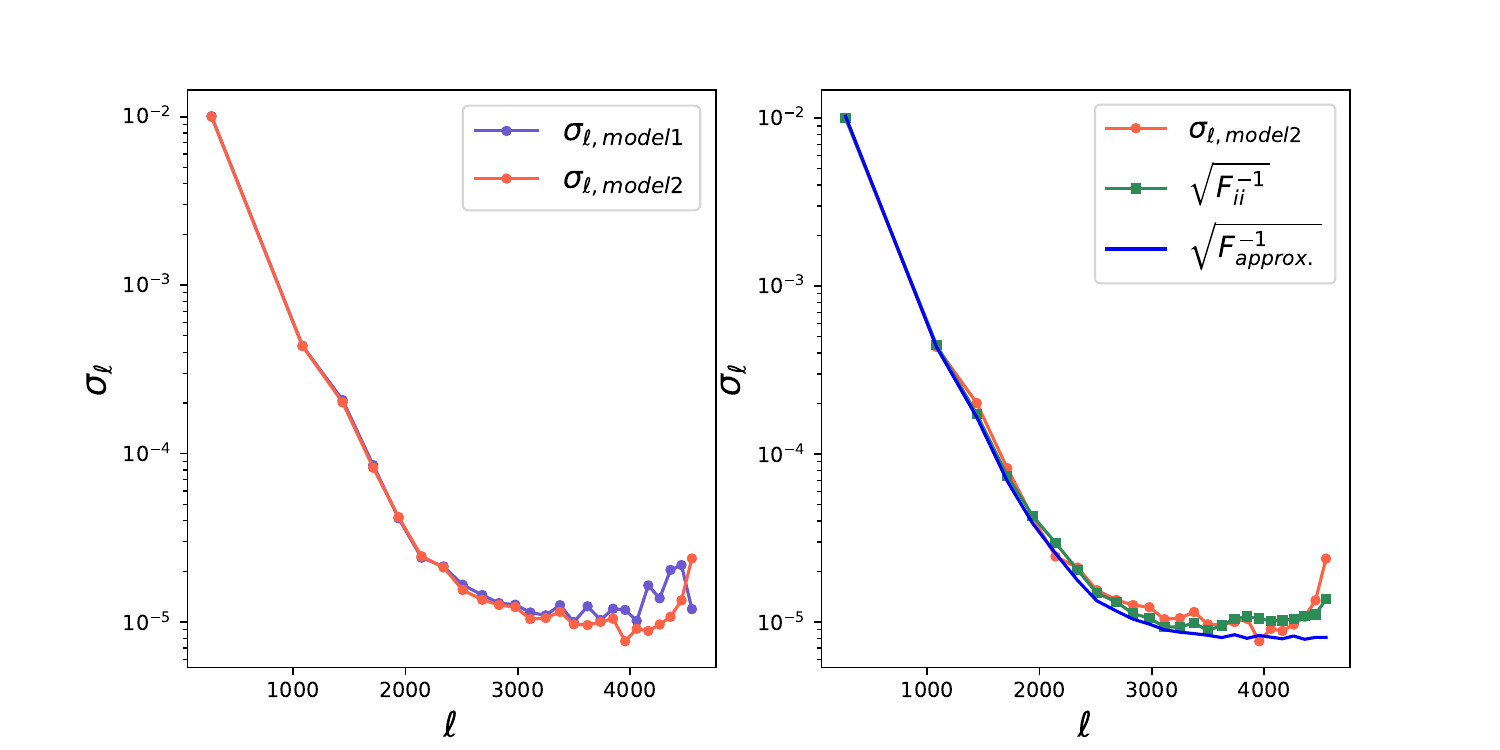}
\caption{Left panel: error of the estimated power spectrum using WienerNet models 1 and 2,  both similar up to scales $\ell=4000$. Right panel:  error of the estimated power spectrum using WienerNet model 2 similar to the square root of the diagonal part of the simulated and approximate inverse Fisher matrix as expected since the inverse of the Fisher matrix is the covariance matrix of the power spectrum.} 
\label{fig:error_homogeneo}
\end{figure}

In the left panel of Figure \ref{fig:error_homogeneo}, the error of the estimated power spectrum using WienerNet models 1 and 2 in the Wiener Filter estimation of the maps is presented. Notably, both errors are similar at least up to scales $\ell = 4000$, indicating that both models are feasible for use in power spectrum estimation.
~

The inverse of the Fisher matrix $F^{-1}$ can be interpreted as an estimate of the covariance matrix of the parameters $\Theta_{\ell}$ for modes that are gaussian distributed. Accordingly, in the right panel of Figure \ref{fig:error_homogeneo} it is shown that the square root of the diagonal elements of the inverse Fisher matrix estimated with \ref{eq:fisher} (green dotted line) is closely similar to the error in the estimation of the \textit{true} power spectrum. The slight difference arises from the fact that the Fisher matrix coincides with the error of the estimation of \textit{fiducial} power spectrum, whereas we are estimating an unknown power spectrum referred to as \textit{true}.

In the case of maps without masks (no mode coupling) the Fisher matrix is diagonal and equal to: 
\begin{equation}
    F_{\ell\ell'} = \frac{2E_\ell^{2}}{K_\ell},
\label{fisher_nomask}
\end{equation}
where $K_\ell$ is the number of modes inside the bin. We can calculate $E_\ell$ with simulations or we can express it in an approximate way knowing that the WF matrix is, in Fourier space, equal to $C_\ell/(C_\ell+n_\ell)$: 
\begin{equation}
    E_\ell = \frac{1}{2} \frac{1}{\Theta_{fid}} \sum_{k_\ell} \frac{C_{k_\ell}}{C_{k_\ell}+n_{k_\ell}}.
\end{equation}

Then to incorporate that the measurements are based on a fraction $f_{sky}$, using the expression \ref{fisher_nomask} for the Fisher matrix without mask, the approximate Fisher matrix is equal to $F_{approx.} = f_{sky} F_{\ell\ell'}$,  which is an ideal expression of the Fisher matrix because it does not consider the details of the mask, just the sky fraction through $f_{sky}$. On the right panel of Figure \ref{fig:error_homogeneo} it is presented both the square root of the diagonal part of the approximate and simulated inverse Fisher matrix (blue solid line and dotted green line, respectively), which are very similar except for the last modes that correspond to the lack of convergence in the simulated Fisher matrix beyond noise-dominated regime.

\section{Results: Inhomogeneous case}
\label{sec:results_inhom}

\subsection{Comparison with \textsc{NIFTy} results}

\begin{figure}
\centering
\includegraphics[width=1\columnwidth]{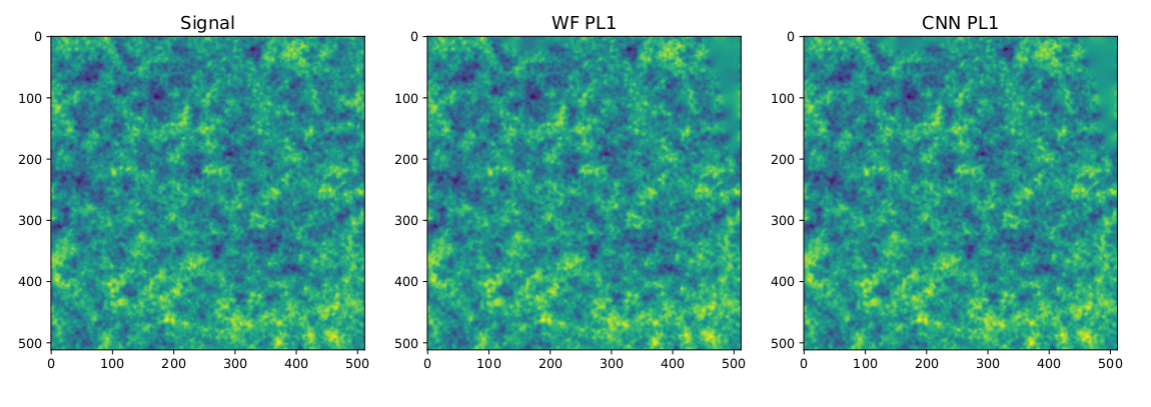}
\caption{
\textcolor{black}{On the left panel we present a signal map, obtained from the power spectrum.  To this map, we applied noise from a noise variance map called ”PL1”, and the mask. Afterwards, we applied the Wiener Filter, both the exact WF and the CNN. The output filtered maps are qualitatively similar, as can be seen in the middle and right panels.}
}
\label{fig:ejemplo_inho}
\end{figure}

Following the same structure as in the homogeneous case, we trained the neural network for maps of $512 \times 512$ pixels and different inhomogeneous noise models, and we found that the best performance that we can obtain is with the same hyperparameters as in model 2 of the homogeneous case, but with 16 number of filters per layer. \textcolor{black}{Figure \ref{fig:ejemplo_inho} compares a map of the signal, generated from the power spectrum, to the filtered map computed with the exact WF or with the trained CNN, after adding inhomogeneous noise, and applying a mask to the original signal map. With this comparison, we can visually  check that both implementations of the Wiener Filter are efficient in reconstructing the signal in unmasked pixels. Furthermore, this shows its capability to  reconstruct large-scale modes even in the masked region.
}

\begin{figure}
\centering
\includegraphics[width=1\columnwidth]{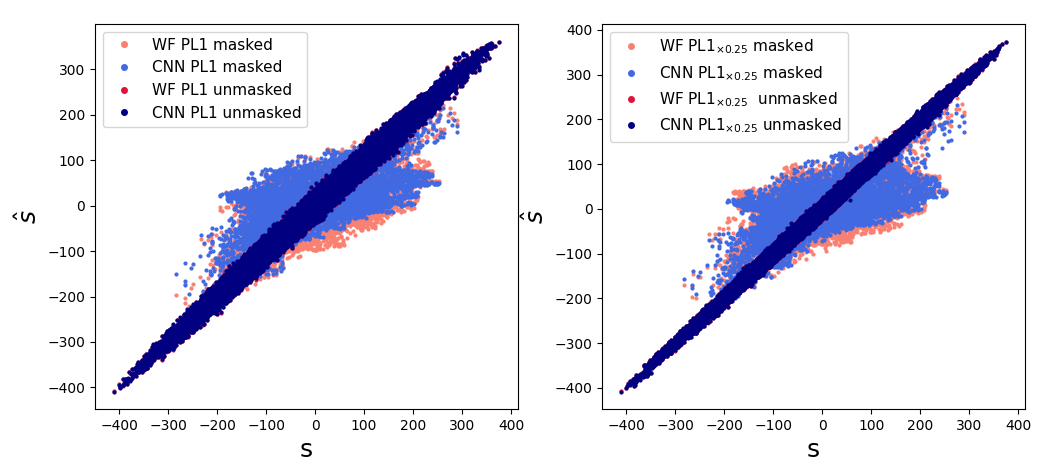}
\caption{Scatter plot of the Wiener Filter pixel predictions ($y$-axis) with respect to the signal map pixels ($x$-axis), the masked pixels are in softer colors with respect to the unmasked pixels. Left panel: scatter plot for PL1 case. Right model: scatter plot for PL1$_{\times 0.25}$ case.}
\label{fig:scatter1}
\end{figure}

Given that the Wiener Filter aims to reconstruct the underlying signal, Figure \ref{fig:scatter1} illustrates a scatter plot representing the relationship between signal map pixels ($x$-axis) and Wiener Filter reconstruction pixels ($y$-axis). The left panel corresponds to the PL1 case, while the right panel represents the PL1$_{\times 0.25}$ scenario. In both cases, it is evident that the neural network's pixel predictions closely align with the unbiased straight line, resembling the results obtained with the \textsc{NIFTy} Wiener Filter. Notably, areas with higher dispersion in the scatter plot coincide with masked regions. It is important to highlight that the scatter plot in the left panel exhibits greater dispersion compared to the right panel due to the higher noise level in PL1 compared to PL1$_{\times 0.25}$.
~

\begin{figure}
\centering
\includegraphics[width=1\columnwidth]{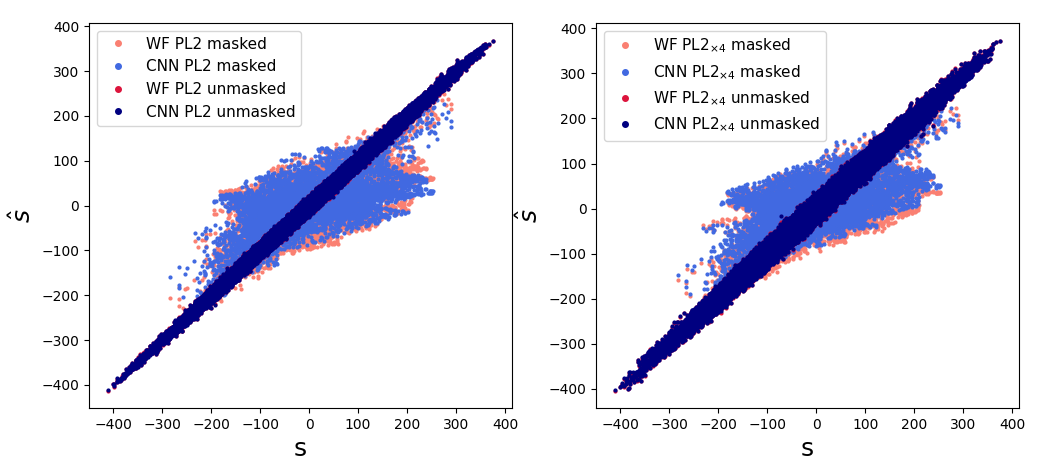}
\caption{Scatter plot of the Wiener Filter pixel predictions ($y$-axis) with respect to the signal map pixels ($x$-axis), the masked pixels are in softer colors with respect to the unmasked pixels. Left panel: scatter plot for PL2 case. Right model: scatter plot for PL2$_{\times 4}$.}
\label{fig:scatter2}
\end{figure}

Similarly, Figure \ref{fig:scatter2} displays a scatter plot of the Wiener Filter reconstruction pixels for PL2 (left panel) and PL2$_{\times4}$ (right panel). The left panel shows less dispersion compared to the right panel, attributed to the lower noise level in PL2 as opposed to PL2$_{\times 4}$.
~

\begin{figure}
\centering
\includegraphics[width=1\columnwidth]{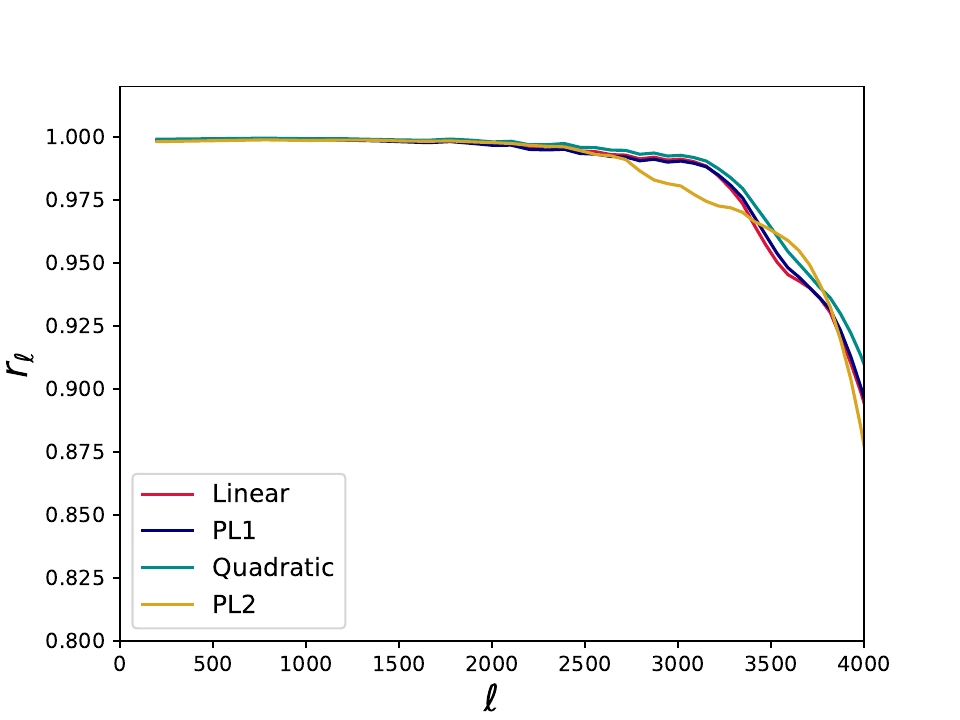}
\caption{Cross correlation coefficient between the neural network model and the exact Wiener Filter. Results for the four noise variance maps specified in section \ref{sec:datasets}. The cross-correlation of PL1, Quadratic, and Linear are close together since the noise levels of these variance maps are similar. PL2 has the lowest average noise but a high degree of inhomogeneity.}
\label{fig:inho_4models}
\end{figure}

Figure \ref{fig:inho_4models} shows the cross-correlation coefficient $r_{\ell}$ between the neural network models and the exact WF result for the quadratic, linear, and Planck samples (every inhomogeneous map variance is a different noise problem to optimize). While a reasonable correlation is observed at scales around 3000, then it begins to decorrelate at higher $\ell$ due to the diminished signal-to-noise ratio.  Also, due to the complexity of the inhomogeneous map variance of PL2 there is a higher decorrelation with respect to the exact WF, than in the other three noise scenarios. The plots are presented up to the scale $\ell=4000$ because the integrated signal-to-noise ratio saturates near those scales.  

\begin{figure}
\centering
\includegraphics[width=1\columnwidth]{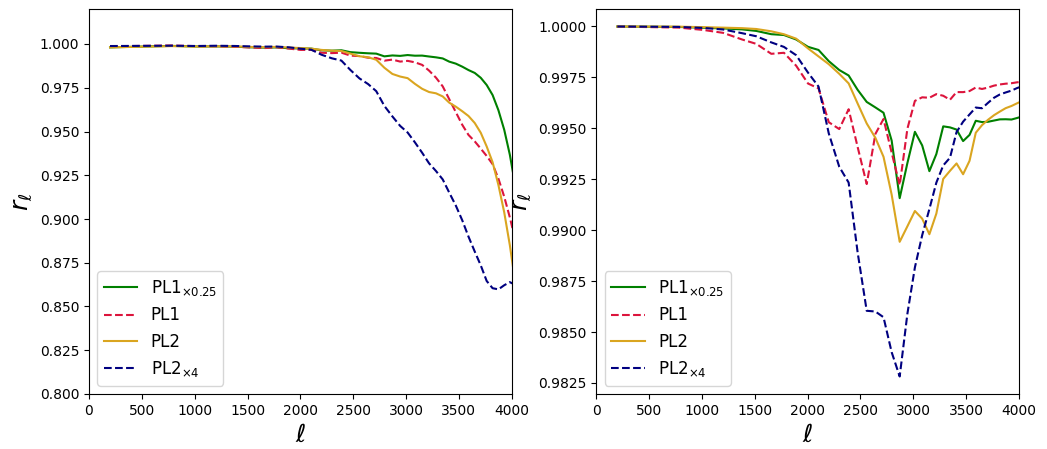}
\caption{ 
Cross-correlation coefficient between the neural network and the exact WF for Planck variance maps and the scaled ones. Left panel: for all the pixels. Right panel: only for the unmasked pixels.
Due to their noise level, PL1 and PL2$_{\times 4}$ begins to decorrrelate at scale $\ell = 2000$, while PL2 and PL1$_{\times 0.25}$ begins to decorrelate approximately at scale $\ell=2500$. However, PL2 and PL2$_{\times 4}$ show a stronger decorrelation with respect to the exact WF. 
Hence, the performance of the neural network is primarily influenced by the complexity of the map variance rather than the noise level.
}
\label{fig:inho_escaled}
\end{figure}

On the other hand, the left panel of Figure \ref{fig:inho_escaled} shows the cross-correlation when the Planck samples 1 and 2 are scaled dividing and multiplying by 4, respectively. The average noise level of PL1 closely matches the average noise level of PL2$_{\times 4}$. Similarly, the average noise level of PL2 closely matches the average noise level of PL1$_{\times 0.25}$, as shown in the power spectrum Figure \ref{espectros_escaled}.
As a result, the correlation pattern between these two pairs exhibits a high degree of similarity. However, PL1 demonstrates stronger correlation compared to PL2$_{\times 4}$, and PL1$_{\times 0.25}$ a stronger correlation compared to PL2. 

PL2 displays a higher degree of inhomogeneity compared to PL1. Even when the map variance is scaled up or down to match the average noise level, the inhomogeneity pattern remains equal. Consequently, the performance of the neural network is primarily influenced by the complexity of the map variance rather than the noise level. This suggests that using the neural network is a viable option for both high and low signal-to-noise ratio problems, as long as the variance map is not excessively inhomogeneous.

As in the homogeneous noise problem, the decorrelation of the neural network concerning the exact Wiener filter results from the masked regions. The right panel of Figure \ref{fig:inho_escaled} shows the cross-correlation coefficients within the mask, where PL1$_{\times 0.25}$ exhibits a stronger correlation than PL2, and PL1 displays a stronger correlation than PL2$_{\times 4}$. The scale at which the correlation begins to decline corresponds to the point where the average noise level intersects the signal, which occurs at scales near $\ell = 2000$ and $\ell = 2500$, respectively. 

\subsection{Implementation of the power spectrum}
\label{sec:powerspectrum_inhom}

For the estimation of the power spectrum in the inhomogeneous case we followed the procedure employed in the homogeneous case using the expressions \ref{blmv} and \ref{eq:fisher} for the noise bias and the Fisher matrix, respectively. We assessed these terms for each noise model scenario (linear, quadratic, and PL1 and PL2).

We computed the noise bias and the Fisher matrix through simulations of a new \textit{fiducial} power spectrum and we estimated a new unknown \textit{true} power spectrum. For the estimation of the Wiener Filter $\bi{\hat{s}}$ in each noise scenario, we utilized the trained models described in the previous section. 

\begin{figure}
\centering
\includegraphics[width=1\columnwidth]{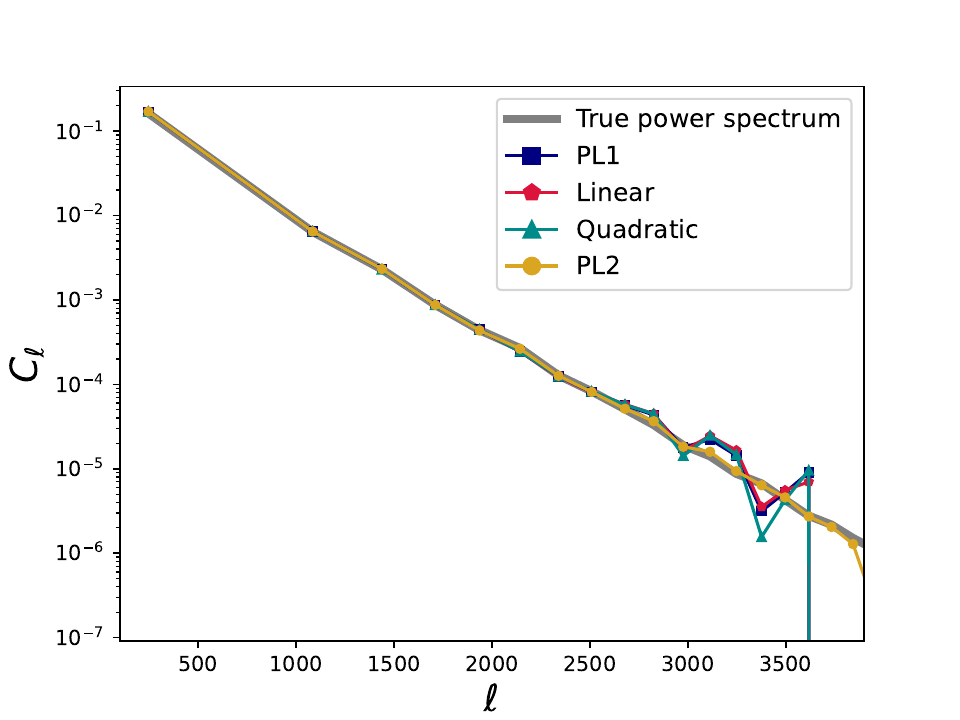}
\caption{Estimation of the true power spectrum for one map in each noise case (linear, quadratic, PL1, and PL2). The estimation in the PL2 case goes to smaller scales than the other noise cases since the average noise level is lower and the scale for noise crossing is greater than for the other ones.}
\label{fig:espectros_modelo}
\end{figure}

Figure \ref{fig:espectros_modelo} illustrates the estimation of the \textit{true} power spectrum for one map in each noise scenario. The estimation of the power spectrum for PL2 is accurate on scales up to $\ell=4000$, while the other three cases are accurate only up to $\ell = 3000$. 

\begin{figure}
\centering
\includegraphics[width=1\columnwidth]{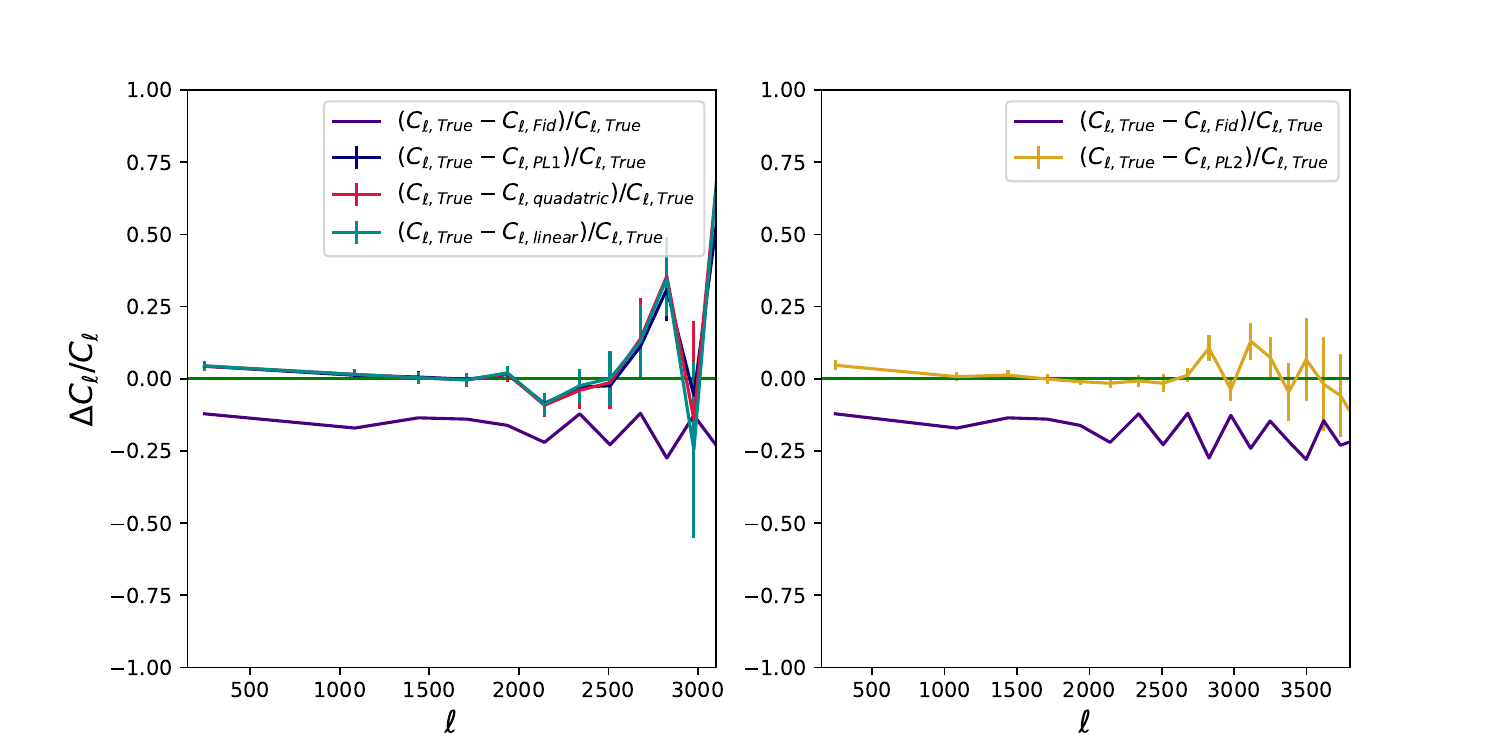}
\caption{Relative difference between the true power spectrum and the estimation of the power spectrum for each noise case. Left panel: cases of linear, quadratic, and PL1, where the error bars become larger beyond scales $\ell = 3000$ and it is not distinguishable from the \textit{fiducial} power spectrum on that scales. Right panel: PL2 case for which the error bars are small enough up to scales near $\ell = 4000$.}
\label{fig:dif_modelos}
\end{figure}

\begin{figure}
\centering
\includegraphics[width=1\columnwidth]{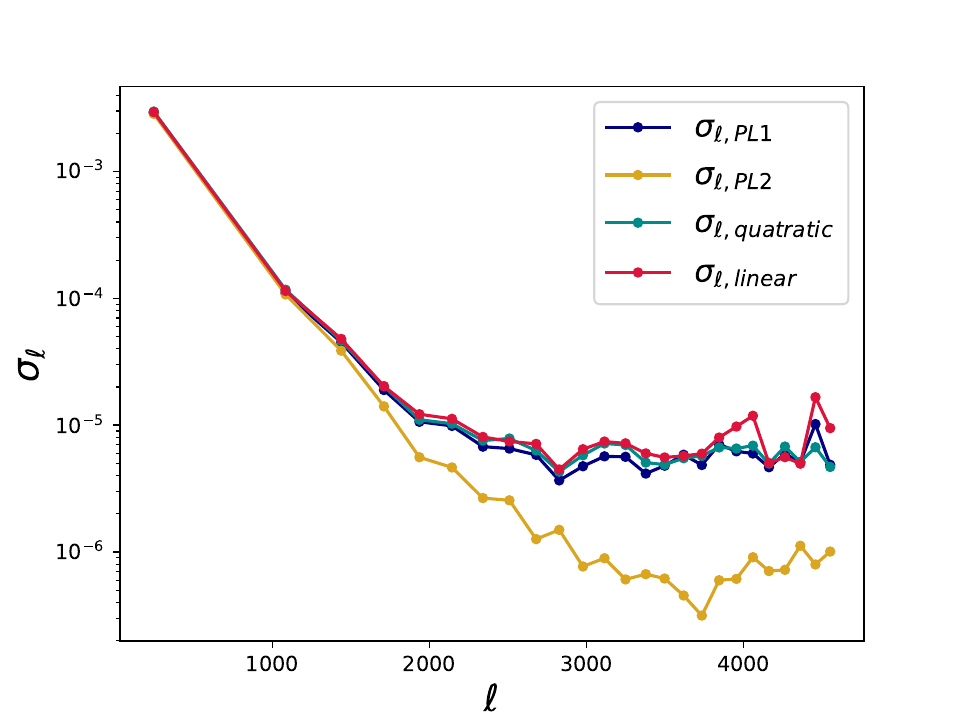}
\caption{Errors in the estimation for the power spectrum for each noise case.  The error of PL2 is lower than the other cases, in agreement with the plots above.}
\label{fig:errores_inho}
\end{figure}

This behavior is better illustrated in Figure \ref{fig:dif_modelos}, where the relative difference between the \textit{true} power spectrum and the estimated power spectrum for PL2 has lower relative error bars up to the scale $\ell=4000$ (right panel), compared to the other cases where the error bars become larger for scales beyond $\ell = 3000$. This discrepancy is attributed to the fact that the average noise level of these three noise scenarios (linear, quadratic, and PL1) is orders of magnitude larger than the signal at those scales, unlike the average noise level of the PL2 case. The lower errors in PL2 compared to the other cases can be observed in Figure \ref{fig:errores_inho}.

\begin{figure}
\centering
\includegraphics[width=1\columnwidth]{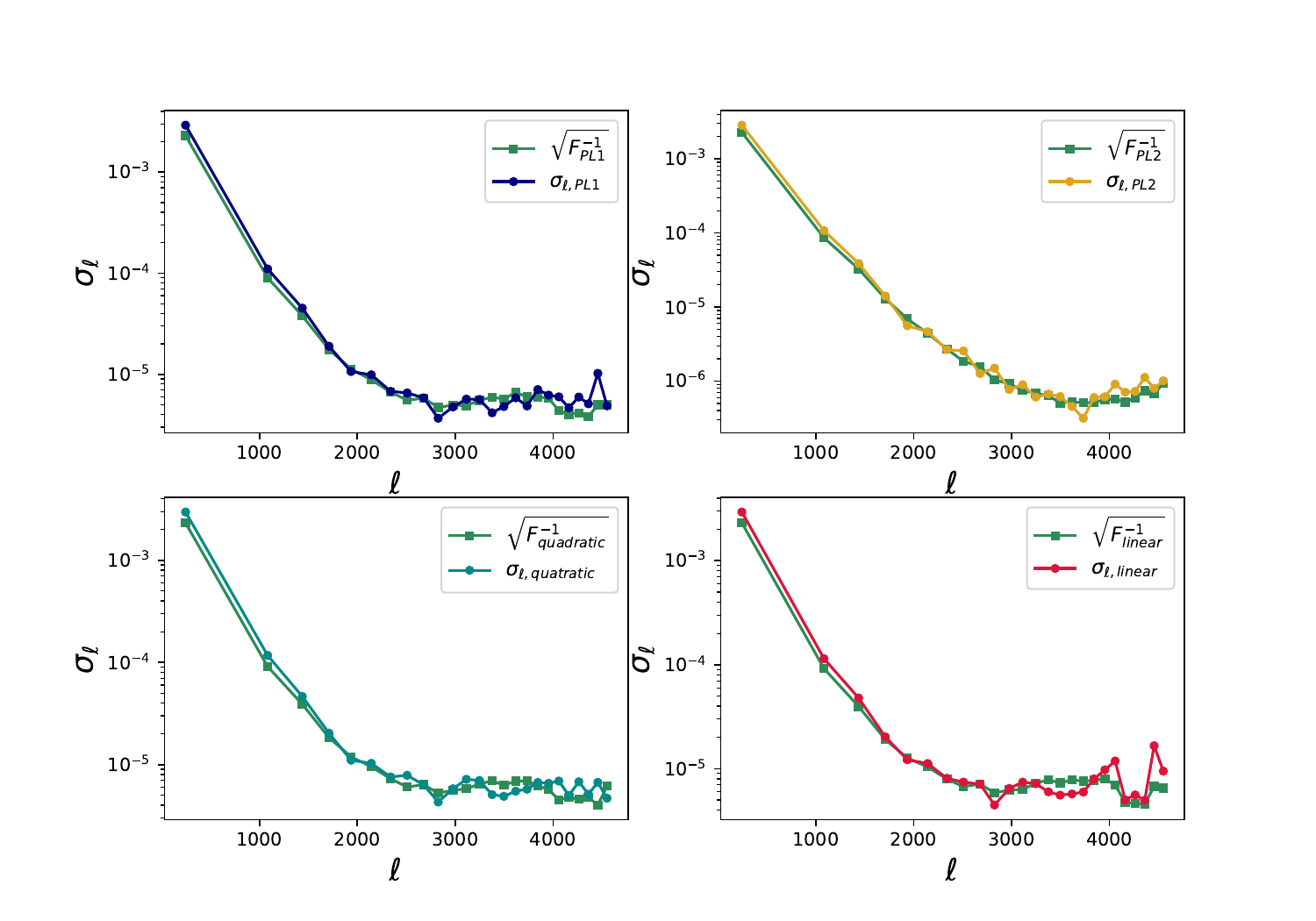}
\caption{ These figures show the square root of the diagonal part of the inverse Fisher matrix and the error of the estimation of the power spectrum, for each noise scenario. The curves are close enough, as expected, since the Fisher matrix is the covariance matrix of the power spectrum.}
\label{fig:fisher_errores}
\end{figure}

Consistent with the theory and the homogeneous results, Figure \ref{fig:fisher_errores} illustrates that the square root of the inverse Fisher matrix for each noise case is similar to the error of the estimation of the \textit{true} power spectrum.

\section{Conclusion}
\label{sec:conclusion}

In this paper, we combine a neural network to apply the Wiener filter to CMB temperature maps with a simulation-based optimal quadratic estimator, to compute the power spectrum of partial and noisy CMB temperature maps, in a reasonable amount of time.

We have demonstrated that the neural network WienerNet, written in an updated version of \textsc{Tensorflow} and \textsc{Keras}, can reproduce the expected results of the exact Wiener Filter for masked CMB maps with homogeneous noise applied.
The reached accuracy depends on the careful selection of hyperparameters, which ensures that optimal performance is achieved.
Furthermore, we have shown that it is feasible to use the trained model for the estimation of the unknown power spectrum of the underlying signal from the maps,  up to modes where the noise is not significantly larger than the signal.

This paper demonstrates the feasibility of estimating the power spectrum through the presented procedure,
which is a simulation-based optimal quadratic estimator,
 because the Wiener Filter prediction performed with the CNN is much faster than the exact method with Conjugate Gradient.
 \textcolor{black}{The CNN takes less than a minute to compute the WF, even for the largest maps. On the other hand, the exact method takes several hours, the time scaling with the size of the map.
  Moreover, the computation time of the CNN prediction does not scale with the multipole of transition from signal-dominated to noise-dominated, while the exact method does. This is because  it takes longer for the exact method in the cases where there are more modes in the signal-dominated region}, as demonstrated in Appendix~\ref{apx1}. 
\textcolor{black}{This leads to an improvement of a factor 1000 or more in time for the largest maps, for which the use of the exact Wiener Filter method is prohibitively.}
 Note that for the calculation of the noise bias term and of the Fisher matrix, that appears in the power spectrum estimator, it is necessary to apply the Wiener Filter to hundreds and thousands of maps (2000 maps for the convergence of the Fisher matrix, and 100 maps for the noise bias), which is impossible to do with the traditional Conjugate Gradient in a reasonable amount of time. 

Besides, in this work, we proposed our own neural network for the implementation of the Wiener Filter for masked CMB temperature maps with inhomogeneous noise, with an additional nonlinear channel to treat the varying variance across pixels in the map. We have explored the problem by proposing different variance maps to account for different levels of complexity in the inhomogeneous noise. 
\textcolor{black}{We shared the trained models in a dedicated repository\footnote{\url{https://github.com/Belencostanza/WF-CNN/tree/main}}.}
We created simplified variance maps from linear and quadratic functions and we also considered two variance maps from the Planck mission to study a more realistic scenario, also scaling them by a factor of 4 to change the average level of noise but maintaining the same inhomogeneity along the map. 

We have shown that the neural network performs satisfactorily across various scenarios, with higher accuracy observed in less inhomogeneous maps, regardless of the average noise level. As it was presented in the cross-correlation coefficient in Figures \ref{fig:inho_4models} and \ref{fig:inho_escaled}, the cases with PL1 and PL1$_{\times 0.25}$ variance maps have better agreement with the exact Wiener Filter results. 

In addition, we have presented the estimation of the power spectrum for these maps with inhomogeneous noise (linear, quadratic, PL1, and PL2) where it is clear that, apart from the optimization, it is possible to recover the underlying power spectrum of the signal 
up to scales for which the noise starts to dominate the power, and beyond.
Despite the increased complexity introduced by the variance map PL2, which poses challenges to the convergence of the neural network toward an optimal solution, it is still feasible to estimate the power spectrum with reasonable precision. This is particularly true for larger modes compared to the cases involving other variance maps. Despite the substantial inhomogeneity in the PL2 map, the average noise level is lower than in other cases. Consequently, the noise spectrum intersects with the signal spectrum at a larger scale, facilitating more accurate power spectrum estimation.

For both homogeneous and inhomogeneous cases, we have shown in Figures \ref{fig:error_homogeneo} and \ref{fig:fisher_errores} that the estimated errors in the power spectrum are similar to the square root of the diagonal part of the inverse Fisher matrix, which is an expected result due to the Fisher matrix definition, showing that the power spectrum procedure was done correctly.

We remark that the convergence becomes more challenging when the variance map is more complex. This effect is more relevant for the efficiency of the training than the noise level of the map.

In future work, we plan to extend this analysis to include polarization CMB maps and implement more complex noise problems, such as correlated noise. We also intend to apply this neural network to real-world CMB experiments, such as the Q~\&~U Bolometric Interferometer for Cosmology, QUBIC~\cite{QUBIC_I},  an experiment that aims at measuring the primordial B-modes polarization of the CMB.


\appendix

\section{Neural Network Architecture}
\label{apx2}

In CNNs, the neurons are connected between layers inside a receptive field of size $f_{h} \times f_{w}$, also called \textit{kernel size}. A given neuron localized on row $i$ and column $j$ will be connected to neurons in the previous layer between rows ($i \times s_{h}$, $i \times s_{h} + f_{h} - 1$) and columns ($j \times s_{w}$, $j \times s_{w} + f_{w} - 1$), where $s_{h}$ and $s_{w}$ are the horizontal and vertical \textit{strides}. The kernel moves through the image depending on the stride $s$, if $s > 1$ the output shape of the layer will be reduced. 

Since several kernels (called \textit{filters}) are applied to the output layer, the latter will have the same number of characteristic maps as the number of filters. If we have an input image of size $[N,M,L]$ where $L$ is the number of channels of the given image, the output shape of the convolutional layer will be $[N',M',F]$, where $F$ is now the number of filters applied to the image and $N', M'$ depend of the chosen stride and the \textit{padding} (indicating that if zeros are added around the image or layer, this also changes the output shape). The output of a neuron in a given convolutional layer is the weighted sum of the inputs: 
\begin{equation}
    z_{i,j,k} = b_{k} + \sum_{\mu = 0}^{f_{h}-1}\sum_{\nu = 0}^{f_{w}-1}\sum_{k' = 0}^{f_{n'}-1} x_{i',j',k'}.w_{\mu, \nu,k',k},
\end{equation}
where
\begin{align}
    i' & = i \: \times \: s_{h} + \mu, \\
    j' & = j \: \times \: s_{w} + \nu.
\end{align}

Then, if we prefer a nonlinear combination of the inputs, it is necessary to apply an activation function to $z_{i,j,k}$. In our case, we applied a ReLU function. On the other hand, the output shape of the convolutional layer will be equal to: 
\begin{equation}
    \frac{input - kernel + 2 \times padding + 1}{stride}
\end{equation}

It is important to highlight that the output of the neural network proposed for the implementation of the Wiener Filter is a linear combination of the input, as the WF theory requires. The outputs of the nonlinear channels added for the treatment in the variance map and the mask, only contribute by multiplying their results to the output layers of the linear channel, but the final output of the neural network only becomes from the linear channel. The choice of adding more nonlinear channels in the inhomogeneous case is motivated by the fact that a more complex configuration of the architecture recovers an optimal solution faster at a reasonable computational cost. 

In Table \ref{table:arq5}, we present the shapes of the convolutional layers in the inhomogenous architecture ($512 \times 512$) for the linear channel, with their corresponding stride and padding. Note that we considered the same value for the horizontal and vertical stride, and we only referred them as \textit{stride}. Also, the input image is extended $512 + 2 \times 128$ in order to recover as output $512$.

\begin{table}[ht!]
\centering
\begin{tabular}{|c c c c c c c|}  
 \hline
  Layers & \textbf{Input} & \textbf{Output} & \textbf{Kernel size} & \textbf{Filters} & \textbf{Stride} & \textbf{Padding}\\ [0.5ex] 
 \hline
 Encoder0 & $768\times768$ & $764\times764$ & $5\times5$ & 16 & 1 & valid\\ [1ex]
 \hline
 Encoder1 & $764\times764$ & $380\times380$ & $5\times5$ & 16 & 2 & valid\\ [1ex]
 \hline
 Encoder2 & $380\times380$ & $188\times188$ & $5\times5$ & 16 & 2 & valid\\ [1ex]
 \hline
 Encoder3 & $188\times188$ & $92\times92$ & $5\times5$ & 16 & 2 & valid\\ [1ex]
 \hline
 Encoder4 & $92\times92$ & $44\times44$ & $5\times5$ & 16 & 2 & valid\\
 [1ex]
 \hline
 Encoder5 & $44\times44$ & $20\times20$ & $5\times5$ & 16 & 2 & valid\\
 [1ex]
 \hline
 Decoder5 & $20\times20$ & $36\times36$ & $5\times5$ & 16 & 2 & valid\\
 [1ex]
 \hline
 Decoder4 & $36\times36$ & $68\times68$ & $5\times5$ & 16 & 2 & valid\\ 
 [1ex]
 \hline
 Decoder3 & $68\times68$ & $132\times132$ & $5\times5$ & 16 & 2 & valid\\ [1ex]
 \hline
 Decoder2 & $132\times132$ & $260\times260$ & $5\times5$ & 16 & 2 & valid\\ [1ex]
 \hline
 Decoder1 & $260\times260$ & $516\times516$ & $5\times5$ & 16 & 2 & valid\\
 [1ex]
 \hline
 Decoder0 & $516\times516$ & $512\times512$ & $5\times5$ & 1 & 1 & valid\\
 [1ex]
 \hline
\end{tabular}
\caption{Neural network architecture for map size $512\times512$ pixels with kernel size $5\times5$.}
\label{table:arq5}
\end{table}

\vspace{10mm}

\section{Efficiency}
\label{apx1}

\begin{figure}[ht!]
\centering
\includegraphics[width=0.9\columnwidth]{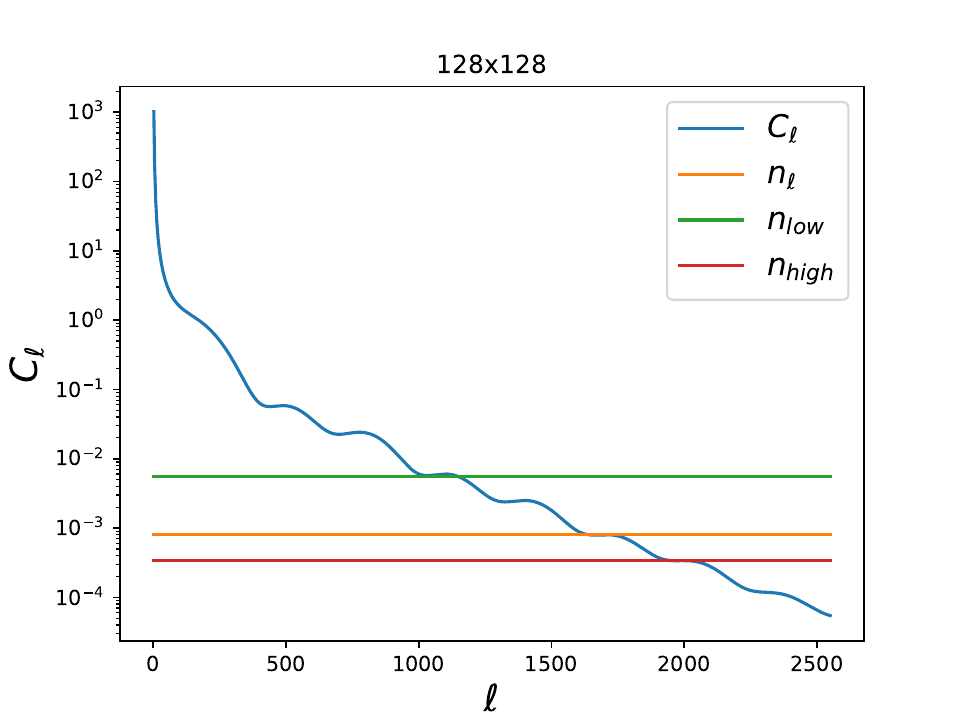}
\caption{Power spectrum of the CMB for $128 \times 128$ maps and three white noise models.}
\label{espectro_128}
\end{figure}

We characterized the noise level using the angular scale, denoted as $\hat \ell$, at which the white noise power spectrum intersects the signal power spectrum, as illustrated in Figure \ref{espectro_128}. Different values of $\hat \ell$ were considered to vary the number of modes with high signal-to-noise ratios, as shown in the figure for maps of $128 \times 128$ pixels.
It is worth noting that the specific value of $\hat \ell$ varies for different map sizes, as indicated in Table~\ref{tab:table_escala}.

\begin{table}[ht!]
\centering
\begin{tabular}{|c c c c c c|} 
\hline
$L_{size}$ & $N_{pix}$ & $\hat{l}$ & $\hat{l}_{low}$ & $\hat{l}_{high}$ & Resolution \\ [0.5ex] 
\hline
10$^\circ$ & 56 & 713 & 504 & 873 & 10.71 arcmin \\ 
10$^\circ$ & 128 & 1629 & 1152 & 1996 & 4.68 arcmin \\
10$^\circ$ & 256 & 3254 & 2304 & 3991 & 2.34 arcmin \\
10$^\circ$ & 512 & 6517 & 4608 & 7982 & 1.17 arcmin \\ [1ex] 
\hline
\end{tabular}
\caption{Simulated noise maps with different number of pixels and three noise levels, used for the study of the efficiency of the CNN, as compared to the CG method.}
\label{tab:table_escala}
\end{table}

\begin{figure}[ht!]
\centering
\includegraphics[width=1.\columnwidth]{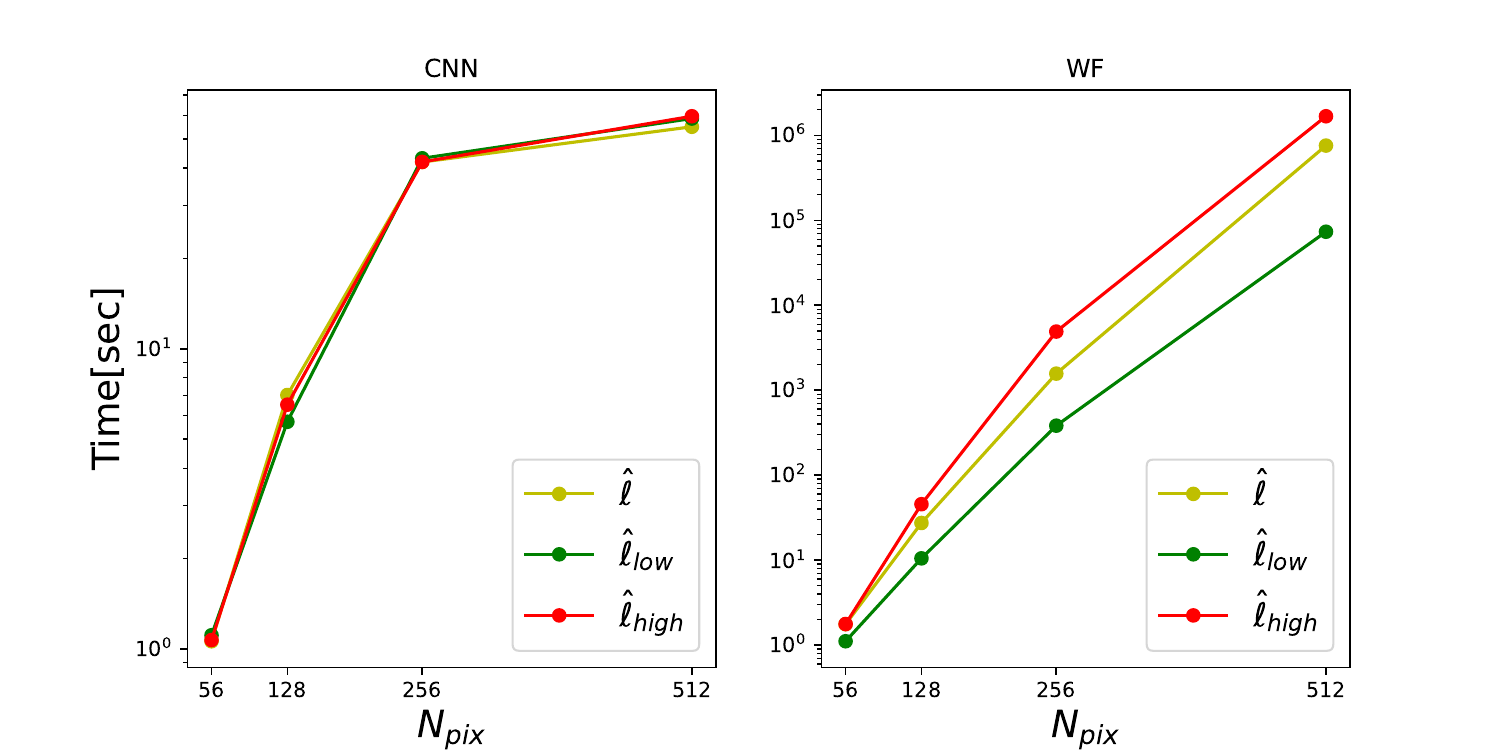}
\caption{Scaling of the computational time required to compute the WF as a function of the number of pixels of the map. Left panel:  using the neural network. Right panel: using the numerical method that accounts for the CG.}
\label{timeWF}
\end{figure}

We trained the neural network for various map sizes and subsequently applied the model to a test set of 300 maps. 
Table \ref{tabla_time_CNN_CG} and Figure \ref{timeWF} shows the computational time, calculated in our CPU Intel core i7 8Th generation, required to compute the Wiener Filter, both for the neural network and the standard method using the CG, implemented in \textsc{NIFTy}. 

The computational time required to compute the Wiener Filter with the neural network is of the order of seconds. It is notable that for the largest map size, $512 \times 512$, it takes less than a minute. These predictions are calculated using the same CPU.

In the case of the CG, for a map size of $56 \times 56$, it takes approximately $1.5$ seconds, for $256 \times 256$, it takes around $2$ hours, and for $512 \times 512$, it takes several days, demonstrating the rapid increase in computational time with map size. Additionally, it can be observed that it takes more time to compute the Wiener Filter with CG in the case of lower noise levels corresponding to the scale $\hat{l}_{high}$, as the signal spectrum predominates over noise, as shown in Figure~\ref{timeWF}. This effect slows down the exact Wiener Filter calculation, a phenomenon not observed in the neural network prediction.


\begin{table}[ht!]
\centering
\begin{tabular}{|l| cccc | cccc | }
\hline
{} & {} &  CNN & {} & {} & {} &  CG  & {} & {} \\
\hline
$l$ & $56$ & $128$ & $256$ & $512$ & $56$ & $128$ & $256$ & $512$  \\
\hline
$\hat{l}$  & 1.06  & 7.01 & 41.9 & 54.9    &  1.76  & 27.4 & 1570 & 759600    \\
$\hat{l}_{low}$ & 1.11 & 5.71 & 43.11 & 58.49    & 1.11 & 10.5 & 384 & 73500  \\
$\hat{l}_{high}$ & 1.07 & 6.51 & 41.93 & 59.52   & 1.77 & 45.6 & 4909 & 1682100 \\
\hline
\end{tabular}
\caption{Computing time, in sec, required to estimate the WF. Left table: using CNN.  Right table: using CG. The value of $l$ characterizes different noise levels for the homogeneous case. }
\label{tabla_time_CNN_CG}
\end{table}


It is worth highlighting that the faster predictions of the Wiener Filter using the neural network, compared to the calculation with the CG method, facilitate the power spectrum procedure. In this procedure, it is necessary to calculate the Fisher matrix and noise bias with simulations. Without the efficiency of WienerNet compared to the exact method, it would be infeasible to perform the further steps developed in this paper.


\vspace{5mm}

\acknowledgments

M.B.C. acknowledges a doctoral fellowship by CONICET, and grant PIP-2876 CONICET.
C.G.S. is supported by grant PIP-2876 CONICET, and grant G175 from UNLP. M.Z. is supported by NSF 2209991 and NSF-BSF 2207583.
The training of the neural networks and the simulations used here were performed at the SNS Supercomputing Center, at the Institute for Advanced Studies, in Princeton.


 \bibliographystyle{JHEP}
 \bibliography{bibliografia_WF.bib}

\providecommand{\href}[2]{#2}\begingroup\raggedright\begin{thebibliography}{10}

\bibitem{Planck2020}
{Planck Collaboration}, N.~{Aghanim}, Y.~{Akrami}, F.~{Arroja}, M.~{Ashdown},
  J.~{Aumont} et~al., \emph{{Planck 2018 results. I. Overview and the
  cosmological legacy of Planck}},
  \href{https://doi.org/10.1051/0004-6361/201833880}{\emph{\aap} {\bfseries
  641} (2020) A1} [\href{https://arxiv.org/abs/1807.06205}{{\ttfamily
  1807.06205}}].

\bibitem{ACT2020a}
S.K.~{Choi}, M.~{Hasselfield}, S.-P.P.~{Ho}, B.~{Koopman}, M.~{Lungu},
  M.H.~{Abitbol} et~al., \emph{{The Atacama Cosmology Telescope: a measurement
  of the Cosmic Microwave Background power spectra at 98 and 150 GHz}},
  \href{https://doi.org/10.1088/1475-7516/2020/12/045}{\emph{\jcap} {\bfseries
  2020} (2020) 045} [\href{https://arxiv.org/abs/2007.07289}{{\ttfamily
  2007.07289}}].

\bibitem{ACT2020b}
S.~{Aiola}, E.~{Calabrese}, L.~{Maurin}, S.~{Naess}, B.L.~{Schmitt},
  M.H.~{Abitbol} et~al., \emph{{The Atacama Cosmology Telescope: DR4 maps and
  cosmological parameters}},
  \href{https://doi.org/10.1088/1475-7516/2020/12/047}{\emph{\jcap} {\bfseries
  2020} (2020) 047} [\href{https://arxiv.org/abs/2007.07288}{{\ttfamily
  2007.07288}}].

\bibitem{SPTPol2018}
J.W.~{Henning}, J.T.~{Sayre}, C.L.~{Reichardt}, P.A.R.~{Ade}, A.J.~{Anderson},
  J.E.~{Austermann} et~al., \emph{{Measurements of the Temperature and E-mode
  Polarization of the CMB from 500 Square Degrees of SPTpol Data}},
  \href{https://doi.org/10.3847/1538-4357/aa9ff4}{\emph{\apj} {\bfseries 852}
  (2018) 97} [\href{https://arxiv.org/abs/1707.09353}{{\ttfamily 1707.09353}}].

\bibitem{1997PhRvD..55.5895T}
M.~{Tegmark}, \emph{{How to measure CMB power spectra without losing
  information}}, \href{https://doi.org/10.1103/PhysRevD.55.5895}{\emph{\prd}
  {\bfseries 55} (1997) 5895}
  [\href{https://arxiv.org/abs/astro-ph/9611174}{{\ttfamily
  astro-ph/9611174}}].

\bibitem{Wiener1949}
N.~Wiener, \emph{Extrapolation, interpolation, and smoothing of stationary time
  series. Vol. 2}, MIT press, Cambridge, MA (1949).

\bibitem{Seljak1998}
U.~{Seljak}, \emph{{Cosmography and Power Spectrum Estimation: A Unified
  Approach}}, \href{https://doi.org/10.1086/306019}{\emph{\apj} {\bfseries 503}
  (1998) 492} [\href{https://arxiv.org/abs/astro-ph/9710269}{{\ttfamily
  astro-ph/9710269}}].

\bibitem{Seljak2017}
U.~{Seljak}, G.~{Aslanyan}, Y.~{Feng} and C.~{Modi}, \emph{{Towards optimal
  extraction of cosmological information from nonlinear data}},
  \href{https://doi.org/10.1088/1475-7516/2017/12/009}{\emph{\jcap} {\bfseries
  2017} (2017) 009} [\href{https://arxiv.org/abs/1706.06645}{{\ttfamily
  1706.06645}}].

\bibitem{10.5555/1403886}
W.H.~Press, S.A.~Teukolsky, W.T.~Vetterling and B.P.~Flannery, \emph{Numerical
  Recipes 3rd Edition: The Art of Scientific Computing}, Cambridge University
  Press, USA, 3~ed. (2007).

\bibitem{2004ApJS..155..227E}
H.K.~{Eriksen}, I.J.~{O'Dwyer}, J.B.~{Jewell}, B.D.~{Wandelt}, D.L.~{Larson},
  K.M.~{G{\'o}rski} et~al., \emph{{Power Spectrum Estimation from
  High-Resolution Maps by Gibbs Sampling}},
  \href{https://doi.org/10.1086/425219}{\emph{\apjs} {\bfseries 155} (2004)
  227} [\href{https://arxiv.org/abs/astro-ph/0407028}{{\ttfamily
  astro-ph/0407028}}].

\bibitem{2004PhRvD..70h3511W}
B.D.~{Wandelt}, D.L.~{Larson} and A.~{Lakshminarayanan}, \emph{{Global, exact
  cosmic microwave background data analysis using Gibbs sampling}},
  \href{https://doi.org/10.1103/PhysRevD.70.083511}{\emph{\prd} {\bfseries 70}
  (2004) 083511} [\href{https://arxiv.org/abs/astro-ph/0310080}{{\ttfamily
  astro-ph/0310080}}].

\bibitem{Smith2007}
K.M.~{Smith}, O.~{Zahn} and O.~{Dor{\'e}}, \emph{{Detection of gravitational
  lensing in the cosmic microwave background}},
  \href{https://doi.org/10.1103/PhysRevD.76.043510}{\emph{\prd} {\bfseries 76}
  (2007) 043510} [\href{https://arxiv.org/abs/0705.3980}{{\ttfamily
  0705.3980}}].

\bibitem{1999ApJ...510..551O}
S.P.~{Oh}, D.N.~{Spergel} and G.~{Hinshaw}, \emph{{An Efficient Technique to
  Determine the Power Spectrum from Cosmic Microwave Background Sky Maps}},
  \href{https://doi.org/10.1086/306629}{\emph{\apj} {\bfseries 510} (1999) 551}
  [\href{https://arxiv.org/abs/astro-ph/9805339}{{\ttfamily
  astro-ph/9805339}}].

\bibitem{Elsner2013}
F.~{Elsner} and B.D.~{Wandelt}, \emph{{Efficient Wiener filtering without
  preconditioning}},
  \href{https://doi.org/10.1051/0004-6361/201220586}{\emph{\aap} {\bfseries
  549} (2013) A111} [\href{https://arxiv.org/abs/1210.4931}{{\ttfamily
  1210.4931}}].

\bibitem{Kodi2017}
D.~{Kodi Ramanah}, G.~{Lavaux} and B.D.~{Wandelt}, \emph{{Wiener filter
  reloaded: fast signal reconstruction without preconditioning}},
  \href{https://doi.org/10.1093/mnras/stx527}{\emph{\mnras} {\bfseries 468}
  (2017) 1782} [\href{https://arxiv.org/abs/1702.08852}{{\ttfamily
  1702.08852}}].

\bibitem{Horowitz2019}
B.~{Horowitz}, U.~{Seljak} and G.~{Aslanyan}, \emph{{Efficient optimal
  reconstruction of linear fields and band-powers from cosmological data}},
  \href{https://doi.org/10.1088/1475-7516/2019/10/035}{\emph{\jcap} {\bfseries
  2019} (2019) 035} [\href{https://arxiv.org/abs/1810.00503}{{\ttfamily
  1810.00503}}].

\bibitem{Munchmeyer2019}
M.~{M{\"u}nchmeyer} and K.M.~{Smith}, \emph{{Fast Wiener filtering of CMB maps
  with Neural Networks}},
  \href{https://doi.org/10.48550/arXiv.1905.05846}{\emph{arXiv e-prints} (2019)
  arXiv:1905.05846} [\href{https://arxiv.org/abs/1905.05846}{{\ttfamily
  1905.05846}}].

\bibitem{2016MNRAS.455.4452A}
J.~{Alsing}, A.~{Heavens}, A.H.~{Jaffe}, A.~{Kiessling}, B.~{Wandelt} and
  T.~{Hoffmann}, \emph{{Hierarchical cosmic shear power spectrum inference}},
  \href{https://doi.org/10.1093/mnras/stv2501}{\emph{\mnras} {\bfseries 455}
  (2016) 4452} [\href{https://arxiv.org/abs/1505.07840}{{\ttfamily
  1505.07840}}].

\bibitem{2001PhRvD..64h3003W}
B.D.~{Wandelt}, E.~{Hivon} and K.M.~{G{\'o}rski}, \emph{{Cosmic microwave
  background anisotropy power spectrum statistics for high precision
  cosmology}}, \href{https://doi.org/10.1103/PhysRevD.64.083003}{\emph{\prd}
  {\bfseries 64} (2001) 083003}
  [\href{https://arxiv.org/abs/astro-ph/0008111}{{\ttfamily
  astro-ph/0008111}}].

\bibitem{QUBIC_universe}
A.~{Mennella}, P.~{Ade}, G.~{Amico}, D.~{Auguste}, J.~{Aumont}, S.~{Banfi}
  et~al., \emph{{QUBIC: Exploring the Primordial Universe with the Q\&U
  Bolometric Interferometer}},
  \href{https://doi.org/10.3390/universe5020042}{\emph{Universe} {\bfseries 5}
  (2019) 42}.

\bibitem{QUBIC_II}
L.~{Mousset}, M.M.~{Gamboa Lerena}, E.S.~{Battistelli}, P.~{de Bernardis},
  P.~{Chanial}, G.~{D'Alessandro} et~al., \emph{{QUBIC II: Spectral polarimetry
  with bolometric interferometry}},
  \href{https://doi.org/10.1088/1475-7516/2022/04/035}{\emph{\jcap} {\bfseries
  2022} (2022) 035} [\href{https://arxiv.org/abs/2010.15119}{{\ttfamily
  2010.15119}}].

\bibitem{Zaroubi1995}
S.~{Zaroubi}, Y.~{Hoffman}, K.B.~{Fisher} and O.~{Lahav}, \emph{{Wiener
  Reconstruction of the Large-Scale Structure}},
  \href{https://doi.org/10.1086/176070}{\emph{\apj} {\bfseries 449} (1995) 446}
  [\href{https://arxiv.org/abs/astro-ph/9410080}{{\ttfamily
  astro-ph/9410080}}].

\bibitem{2013A&A...554A..26S}
M.~{Selig}, M.R.~{Bell}, H.~{Junklewitz}, N.~{Oppermann}, M.~{Reinecke},
  M.~{Greiner} et~al., \emph{{NIFTY - Numerical Information Field Theory. A
  versatile PYTHON library for signal inference}},
  \href{https://doi.org/10.1051/0004-6361/201321236}{\emph{\aap} {\bfseries
  554} (2013) A26} [\href{https://arxiv.org/abs/1301.4499}{{\ttfamily
  1301.4499}}].

\bibitem{chen_2005}
K.~Chen, \emph{Matrix Preconditioning Techniques and Applications}, Cambridge
  Monographs on Applied and Computational Mathematics, Cambridge University
  Press (2005),
  \href{https://doi.org/10.1017/CBO9780511543258}{10.1017/CBO9780511543258}.

\bibitem{Chollet2017}
F.~Chollet, \emph{Deep Learning with Python}, Manning Publications, New York,
  NY (2017).

\bibitem{UNet}
O.~{Ronneberger}, P.~{Fischer} and T.~{Brox}, \emph{{U-Net: Convolutional
  Networks for Biomedical Image Segmentation}},
  \href{https://doi.org/10.48550/arXiv.1505.04597}{\emph{arXiv e-prints} (2015)
  arXiv:1505.04597} [\href{https://arxiv.org/abs/1505.04597}{{\ttfamily
  1505.04597}}].

\bibitem{Goodfellow2016}
I.~Goodfellow, Y.~Bengio and A.~Courville, \emph{Deep Learning}, MIT Press
  (2016).

\bibitem{Belen_BAAA2023}
M.B.~{Costanza}, C.G.~{Sc{\'o}ccola} and M.~{Zaldarriaga}, \emph{{Wiener Filter
  for cosmic microwave background maps using neural networks}}, {\emph{Boletin
  de la Asociacion Argentina de Astronomia La Plata Argentina} {\bfseries 64}
  (2023) 193}.

\bibitem{CAMB}
A.~{Lewis}, A.~{Challinor} and A.~{Lasenby}, \emph{{Efficient Computation of
  Cosmic Microwave Background Anisotropies in Closed Friedmann-Robertson-Walker
  Models}}, \href{https://doi.org/10.1086/309179}{\emph{\apj} {\bfseries 538}
  (2000) 473} [\href{https://arxiv.org/abs/astro-ph/9911177}{{\ttfamily
  astro-ph/9911177}}].

\bibitem{planck2013}
{Planck Collaboration}, P.A.R.~{Ade}, N.~{Aghanim}, C.~{Armitage-Caplan},
  M.~{Arnaud}, M.~{Ashdown} et~al., \emph{{Planck 2013 results. XVI.
  Cosmological parameters}},
  \href{https://doi.org/10.1051/0004-6361/201321591}{\emph{\aap} {\bfseries
  571} (2014) A16} [\href{https://arxiv.org/abs/1303.5076}{{\ttfamily
  1303.5076}}].

\bibitem{2014arXiv1412.6980K}
D.P.~{Kingma} and J.~{Ba}, \emph{{Adam: A Method for Stochastic Optimization}},
  \href{https://doi.org/10.48550/arXiv.1412.6980}{\emph{arXiv e-prints} (2014)
  arXiv:1412.6980} [\href{https://arxiv.org/abs/1412.6980}{{\ttfamily
  1412.6980}}].

\bibitem{article}
X.~Glorot and Y.~Bengio, \emph{Understanding the difficulty of training deep
  feedforward neural networks}, {\emph{Journal of Machine Learning Research -
  Proceedings Track} {\bfseries 9} (2010) 249}.

\bibitem{QUBIC_I}
J.C.~{Hamilton}, L.~{Mousset}, E.S.~{Battistelli}, P.~{de Bernardis},
  M.A.~{Bigot-Sazy}, P.~{Chanial} et~al., \emph{{QUBIC I: Overview and science
  program}}, \href{https://doi.org/10.1088/1475-7516/2022/04/034}{\emph{\jcap}
  {\bfseries 2022} (2022) 034}
  [\href{https://arxiv.org/abs/2011.02213}{{\ttfamily 2011.02213}}].

\end{thebibliography}\endgroup

\end{document}